\documentclass{article}

\usepackage{arxiv}
\usepackage[utf8]{inputenc} 
\usepackage[T1]{fontenc}    
\usepackage{hyperref}       
\usepackage{url}            
\usepackage{booktabs}       
\usepackage{amsfonts}       
\usepackage{nicefrac}       
\usepackage{microtype}      
\usepackage{lipsum}
\usepackage{graphicx}
\usepackage{notoccite}
\usepackage{subcaption}
\usepackage{array}
\usepackage{multirow} 
\usepackage{adjustbox}
\usepackage{wrapfig}
\usepackage{mathpazo} 
\usepackage[hang,flushmargin]{footmisc}
\usepackage[backend=biber, style=ieee]{biblatex}
\usepackage{amsmath}
\usepackage{makecell}
\usepackage{units}
\usepackage{verbatim}
\usepackage{longtable}
\usepackage{appendix}

\usepackage[font={small}]{caption} 
\hypersetup{hidelinks} 

\graphicspath{ {./figures/} }

\title{Auditing Google's Search Algorithm: Measuring News Diversity Across Brazil, the UK, and the US}

\author{ \href{https://orcid.org/0009-0005-0323-8326}{\includegraphics[scale=0.06]{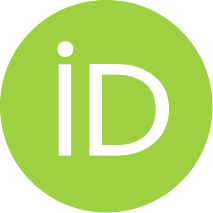}\hspace{1mm}Raphael Hernandes}\\
	Leverhulme Centre for the Future of Intelligence\\University of Cambridge\\
	\texttt{rhh43@cam.ac.uk}
    \And
    \href{https://orcid.org/0000-0002-5130-2258}{\includegraphics[scale=0.06]{orcid.pdf}\hspace{1mm}Giulio Corsi}\\ 
    Leverhulme Centre for the Future of Intelligence\\University of Cambridge\\
    \texttt{gc540@cam.ac.uk} 
}

\hypersetup{
pdftitle={Auditing Google's Search Algorithm: Measuring News Diversity Across Brazil, the UK, and the US},
pdfsubject={ethics of ai},
pdfauthor={Raphael Hernandes, Giulio Corsi},
pdfkeywords={news diversity, artificial intelligence, Google, news exposure, search engine, algorithm audit},
}

\begin{document}

\maketitle

\vspace{10pt}

\begin{abstract}
This study examines the influence of Google's search algorithm on news diversity by analyzing search results in Brazil, the UK, and the US. It explores how Google's system preferentially favors a limited number of news outlets. Utilizing algorithm auditing techniques, the research measures source concentration with the Herfindahl-Hirschman Index (HHI) and Gini coefficient, revealing significant concentration trends. The study underscores the importance of conducting horizontal analyses across multiple search queries, as focusing solely on individual results pages may obscure these patterns. Factors such as popularity, political bias, and recency were evaluated for their impact on news rankings. Findings indicate a slight leftward bias in search outcomes and a preference for popular, often national outlets. This bias, combined with a tendency to prioritize recent content, suggests that Google's algorithm may reinforce existing media inequalities. By analyzing the largest dataset to date---221,863 search results---this research provides comprehensive, longitudinal insights into how algorithms shape public access to diverse news sources.
\end{abstract}


\vspace{10pt}

\section{Introduction}

News diversity is crucial for the health of democratic processes. It allows for a wide range of opinions to be heard, which is essential for a well-informed public, and aids the reception of diverse ideas~\parencite{napoliWhatMediaPolicy2023}, \parencite{vanderwurffAudiencesReceiveDiverse2011a}. In the digital age, part of that diversity comes from having an ecosystem of multiple outlets with different ideologies and framings to expose people to various viewpoints. However, this will only work if the content actually gets to readers. News distribution, therefore, plays a crucial role in this process~\parencite[248,254-257]{napoliExposureDiversityReconsidered2011}.

Algorithms increasingly mediate the distribution of news. The digital age brought an uptick in the offering of information, and these systems are intermediaries in content discovery. Their behavior can have a significant impact on what people see~\parencite[177-203]{diakopoulosAutomatingNewsHow2019}, \parencite[]{thurmanAlgorithmsAutomationNews2019}, \parencite[1683-1685]{evansGoogleNewsMachine2023}. Given their important role, it is necessary to investigate how they might affect news distribution.

The importance of ensuring news diversity in these systems is twofold. It fosters a healthy news ecosystem by exposing readers to a variety of content, and the web traffic generated from the exposure brings economic benefits to publishers~\parencite[248]{napoliExposureDiversityReconsidered2011}, \parencite[13]{trielliSearchNewsCurator2019}.

Search engines are crucial and act as gatekeepers in this process, as most readers reach news websites through them~\parencite{GlobalAudienceInsights2024}, \parencite[]{thurmanAlgorithmsAutomationNews2019}, \parencite[50]{paulussenNewsValuesAudienceOriented2021}, \parencite[210]{bozdagBiasAlgorithmicFiltering2013a}. Google is an undisputed leader in this field, with over 90\% of market share worldwide~\parencite{mattg.southernGoogleSearchEngine2024}, \parencite{SearchEngineMarket2024}.

In the literature regarding diversity in the media, this is related to the idea of \textit{exposure diversity}: the variety of content and viewpoints that readers access, which goes through exposing audiences to different sources~\parencite[149]{moeOperationalizingExposureDiversity2021}, \parencite[24]{napoliDeconstructingDiversityPrinciple1999}, \parencite[330]{vanderwurffAudiencesReceiveDiverse2011a}. By displaying news in its results, Google diverts readers' attention to the outlets and articles it curates~\parencite[3]{trielliSearchNewsCurator2019}.

While previous research has identified biases in these search engines~\parencite{evansGoogleNewsMachine2023}, \parencite{hannakMeasuringPersonalizationWeb2017}, \parencite{trielliSearchNewsCurator2019}, these analyses often have limitations in scope, such as focusing solely on the US or English-language content, or examining a narrow set of queries. This paper extends the literature by investigating the presence of biases in the news distribution in Google Search through algorithm auditing techniques in Brazil, the UK, and the US~\parencite[6]{metaxaAuditingAlgorithmsUnderstanding2021}, using metrics that have been explored in this context in the past (such as popularity and political bias) and incorporating new ones: information quality and the number of articles each outlet published on a given topic. By extracting data from the "News" tab in Google Search over 21 days, it analyzed 143,976 results related to 2,298 search queries and 4,296 outlets. Key parts of the findings were validated on an additional dataset collected over 12 days, with 77,887 results. The analysis shows high market concentration in Google Search's algorithm outputs, along with biases relating to the sources' popularity and political stance, and the content's recency.

\section{Background}\label{sec:background}

\subsection{News Diversity and Algorithms Mediating News Consumption}\label{sec:algomediators}

Diversity is widely regarded as desirable in news ecosystems and crucial for democratic processes~\parencite[2]{mcquailDiversityMediaPolicy1983}, \parencite[]{napoliWhatMediaPolicy2023}, \parencite[]{vanderwurffAudiencesReceiveDiverse2011a}, \parencite[1]{napoliDeconstructingDiversityPrinciple1999}. Its components include diversity of sources (news organizations), content (viewpoints and ideas) and exposure (consumption)~\parencite[247-248]{napoliExposureDiversityReconsidered2011}, \parencite[10]{napoliDeconstructingDiversityPrinciple1999}. Ensuring diversity is one goal that drives policy in this area, and in recent years these have turned their attention to \textit{exposure diversity} ---the diversity of content and viewpoints that readers actually access~\parencite[149]{moeOperationalizingExposureDiversity2021}, \parencite[24]{napoliDeconstructingDiversityPrinciple1999}.

The distinction between internal diversity (within media outlets) and external diversity (between media outlets) becomes particularly relevant in this context as, with more niche and partizan media appearing, the latter helps ensure balance~\parencite[330]{vanderwurffAudiencesReceiveDiverse2011a}, \parencite[251]{napoliExposureDiversityReconsidered2011}, \parencite[179]{stackhouseNarrowcastingHowMedia2016}. In a scenario where audiences that were once choosing from a small number of publications then have an excess of choices for content~\parencite[251]{napoliExposureDiversityReconsidered2011}, \parencite[179]{stackhouseNarrowcastingHowMedia2016}, the abundance of content can paradoxically lead to polarization if users expose themselves to like-minded information instead of diverse sources~\parencite[330]{vanderwurffAudiencesReceiveDiverse2011a}, \parencite[252]{napoliExposureDiversityReconsidered2011}. Therefore, ensuring that diverse content gets to people is crucial for the diversity of the news ecosystem to be effective.

Readers increasingly rely on tech platforms like Facebook and Google to access news content instead of going directly to news websites. These "side-door" routes are particularly popular with younger audiences and are becoming more common over time~\parencite[10]{newmanReutersInstituteDigital2023}, with search engines accounting for 17\% to 41\% of traffic to news websites, depending on the country~\parencite{GlobalAudienceInsights2024}. 

While algorithms, acting as gatekeepers, help users sort through the information overload common in the digital age, they are not free from bias, despite what some ideas of having these allegedly neutral alternatives might suggest~\parencite[209-214]{bozdagBiasAlgorithmicFiltering2013a}, \parencite[182]{diakopoulosAutomatingNewsHow2019}, \parencite[]{thurmanAlgorithmsAutomationNews2019}, \parencite[45,49--50]{paulussenNewsValuesAudienceOriented2021}. Algorithmic biases might be hidden in systems even for those who develop it~\parencite[331]{batyafriedmanBiasComputerSystems1996}, \parencite[11-12]{philipbreyDisclosiveComputerEthics2000}, \parencite[]{tavaniSearchEnginesEthics2020}. These biases can be technical, such as the criteria for selecting and ranking news, or stem from the humans who design them, reflecting their values and beliefs~\parencite{bozdagBiasAlgorithmicFiltering2013a}.

Google's search engine has shown multiple instances of these biases over the years, including in news distribution~\parencite[18]{ericulkenQuestionBalanceAre05}, \parencite[]{ericulkenNontraditionalSourcesCloud2005}, \parencite[1692-1693]{evansGoogleNewsMachine2023}. It uses a complex algorithm to sort through the vast amount of information on the web and present its pick of the most relevant results to the user. Examples of biases on its technical side include not getting information from every existing website and prioritizing older web domains over newly created ones. On a company side, Google might remove content following requests from governments~\parencite[215-217]{bozdagBiasAlgorithmicFiltering2013a}, or give more space to services owned by Google than their competitors~\parencite{wrightDefiningMeasuringSearch2011}.

In a context that relates to diversity of viewpoints, research has found that search engines may display political biases~\parencite{kulshresthaSearchBiasQuantification2019}. These biases might reflect the input (when the system indexes skewed data, which is reflected in the output). However, they may also relate to the ranking mechanisms surfacing more content that aligns with a political ideology. Google's results, for example, were shown to match the inclinations of the entity represented in the searched keywords, meaning queries about Democrat US politicians generally had left-skewed results~\parencite[215]{kulshresthaSearchBiasQuantification2019}.

The media environment is shifting from broadcasting, where a single message is sent to a large audience, to narrowcasting and hyper-personalization, where content is tailored to individuals, which can lead to a decrease in exposure to diverse viewpoints and ideological segregation~\parencite[798-799]{metzgerBroadcastingNarrowcastingMass2017}. There is an abundance of information and viewpoints available to users at any time if they are part of the 57\% of the world population with internet access on their phones~\parencite[4]{matthewshanahanStateMobileInternet2023}. The question is, thus, to ensure that these people are not exposed to only a small and uniform chunk of this content~\parencite[802]{metzgerBroadcastingNarrowcastingMass2017} \parencite[179-180]{stackhouseNarrowcastingHowMedia2016}.

However, companies that create platforms and algorithms that mediate the information environment aim to maximize profits by increasing consumption rather than ensuring exposure diversity. As suggested by the filter-bubble hypothesis, this misalignment can filter out content that might be less interesting to consumers regardless of their journalistic importance~\parencite[2]{haimBurstFilterBubble2018}. While empirical evidence might challenge the effects of echo chambers and shows that concerns around filter-bubble effects could be exaggerated~\parencite[]{nyhanLikemindedSourcesFacebook2023} \parencite[18]{rossarguedasEchoChambersFilter2022}, their mitigation hinges on users accessing diverse media. Lack of diversity in algorithms that curate news should, therefore, still be a point of concern~\parencite[3]{trielliSearchNewsCurator2019}[22-23]{rottgerInformationEnvironmentIts2020}. In a recent internal data leak from Google, confirmed as authentic by the company, suggests that the search engine takes user engagement into account when ranking results~\parencite{satoGoogleConfirmsLeaked2024}, \parencite{satoBiggestFindingsGoogle2024}.

Google Search is met with great user trust, which has been empirically demonstrated~\parencite{panGoogleWeTrust2007}. This can be dangerous. Switching the order of search results, for example, can shift the voting preferences of undecided voters by 20\% or more~\parencite{epsteinSearchEngineManipulation2015}. And Google's influence in news diffusion is exacerbated by a matter of scale. According to Similar Web, a service that specializes in website audience, the US-based news outlet with the largest audience, \texttt{nytimes.com}, had 633.9 million worldwide visits\footnote{Number of times a user enters a website} in April/2024. \texttt{Google.com} had 70.8 billion, excluding subdomains (such as the email service \textit{Gmail}) and local domains ---the Brazilian version, \texttt{Google.com.br}, for example, had 503.6 million visits~\parencite{WebsitePerformance2024}, \parencite{majidTop50Biggest2024}.

\subsubsection{Google's Algorithms and the Media Market} 

As one of the most important sources of traffic to news websites, Google plays a crucial role in the business model of digital news outlets, which largely relies on selling subscriptions and advertisement-based revenue~\parencite{charlottetobittGlobalNewsIndustry2024}, \parencite{kristenhareHereHowSubscriber2018}. 

Google's relationship with media across the world has multiple aspects. In some, the tech giant supports content producers, which, in turn, ensure Google has information to show to its users. This support comes in the form of web traffic but also partnerships and funding for news initiatives~\parencite[116-118]{nielsenOurFuturesAre2022}, \parencite[162,166,171]{nielsenPlatformPower2022}, \parencite[]{NewsEquityFund2022}, \parencite[]{DigitalNewsInnovation}. On the other hand, this relationship also comes with tensions. Changes to these systems can lead to declines in traffic, and optimizing for performance on such algorithms affects decisions about what and when to post, creating an uncertain environment for content producers~\parencite[179-180,186]{diakopoulosAutomatingNewsHow2019}, \parencite[49]{paulussenNewsValuesAudienceOriented2021}, \parencite[]{munoriyarwaPhilanthrocapitalismGoogleNews2024}.

Previous studies have identified a biased selection of news sources in Google's systems. Analyzing the news aggregator Google News in Germany, research found that \textit{Focus Online} and \textit{Die Welt} had substantial shares of exposures, which was disproportionate considering their actual reach in the country. At the same time, more popular outlets seemed underrepresented~\parencite[9]{haimBurstFilterBubble2018}. High source concentration across different search queries has been found in the US market along with a slight political skewness towards the left compared to a baseline and a favoring of recent articles~\parencite{trielliSearchNewsCurator2019}. In another investigation of two keywords in the US, market concentration was also found with five sources having 93\% of the results for the first keyword and ten sources corresponding to 80\% of the results for the second one, gathering data every minute for 24 hours\parencite[177-179]{diakopoulosAutomatingNewsHow2019} These patterns imply an uneven capture of economic benefits by players in this industry~\parencite[13]{trielliSearchNewsCurator2019}.

\section{Methodology}\label{sec:methods}

\begin{figure}[htb]
    \centering
    \begin{minipage}[b]{0.45\linewidth}
        \centering
        \includegraphics[width=\linewidth]{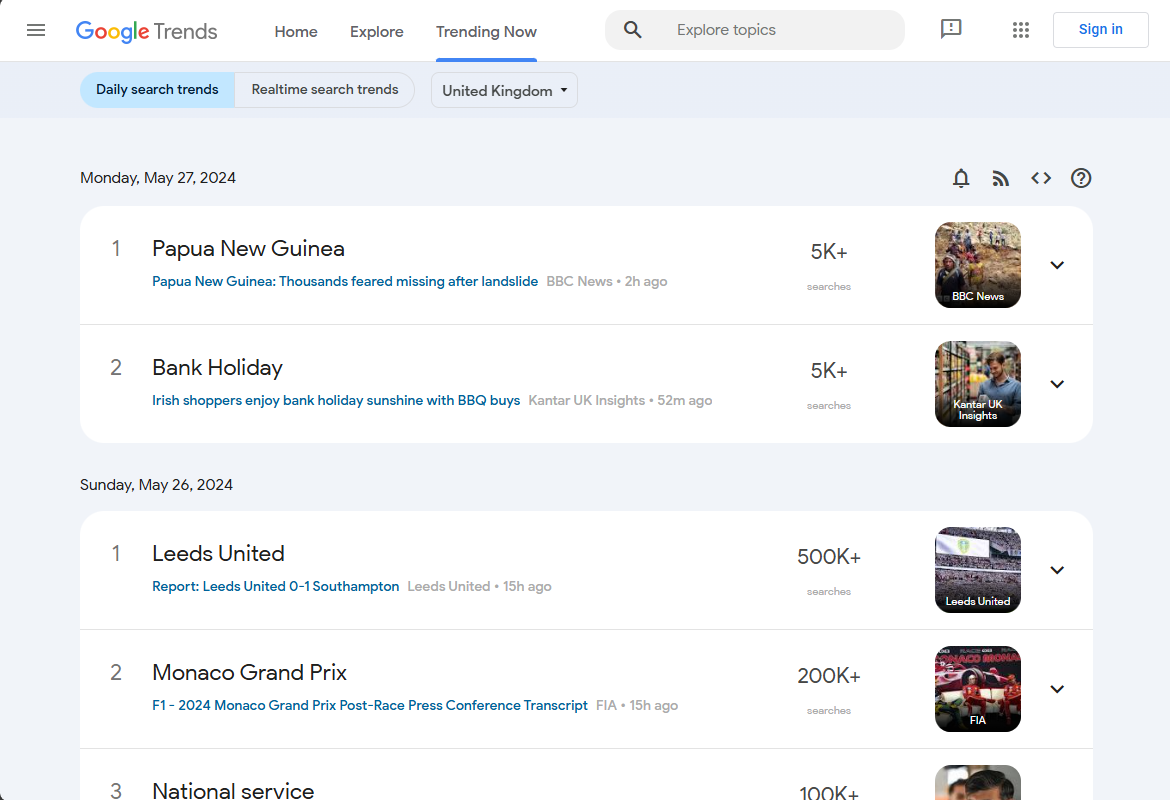}
        \caption*{1}
    \end{minipage}
    \hspace{0.05\linewidth}
    \begin{minipage}[b]{0.45\linewidth}
        \centering
        \includegraphics[width=\linewidth]{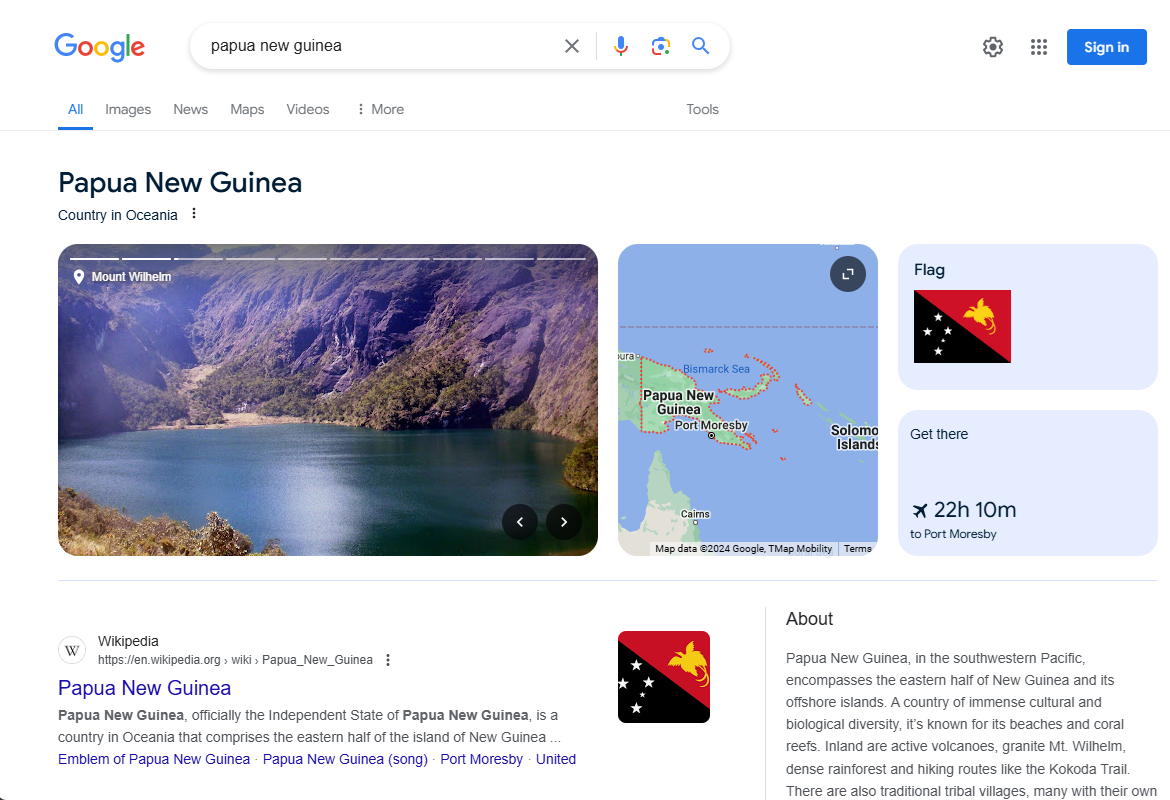}
        \caption*{2}
    \end{minipage}
    \vfill
    \begin{minipage}[b]{0.45\linewidth}
        \centering
        \includegraphics[width=\linewidth]{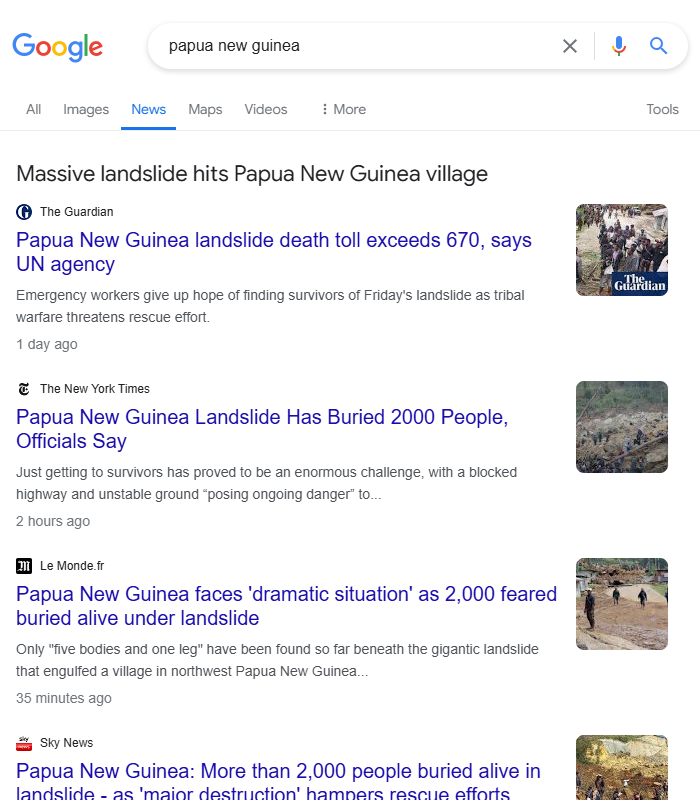}
        \caption*{3}
    \end{minipage}
    \hspace{0.05\linewidth}
    \begin{minipage}[b]{0.45\linewidth}
        \centering
        \includegraphics[width=\linewidth]{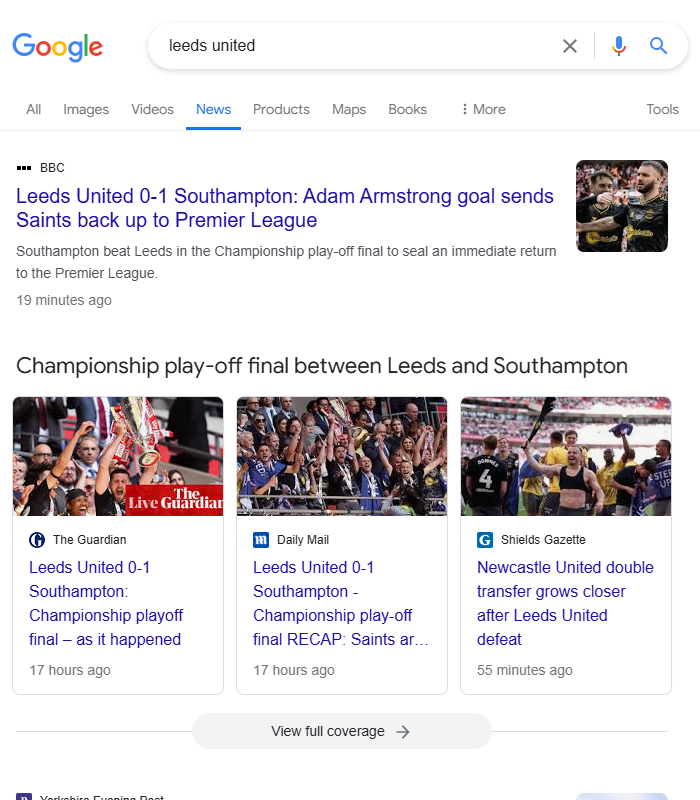}
        \caption*{4}
    \end{minipage}
    \caption{The relevant Google pages mentioned in this analysis. Keywords identified in Google Trends (picture 1) are queried in Search (2). Clicking "News" below the search box leads to the news tab (3). The layout for the results in the news tab varies, all were captured (4).}
\end{figure}

\subsection{Challenges to Evaluating Bias in Search Engines}\label{sec:challenges}

Data from Google's search results is not readily available to researchers, and there is no way to determine the exact results consumed by users on the platform. Algorithm auditing techniques allow third parties to test deficiencies in artificial intelligence systems without access to their internal workings~\parencite[6]{metaxaAuditingAlgorithmsUnderstanding2021}. Scraping Google's results can provide an approximation of what the public receives~\parencite[5]{trielliSearchNewsCurator2019}, \parencite[5]{RecommendationsYoutubeCase2022}, \parencite[22]{metaxaAuditingAlgorithmsUnderstanding2021}, \parencite[]{diakopoulosAuditingHumanMachineCommunication2021}. Google's content is personalized, so the results page that two users get for the same query might be different~\parencite{hannakMeasuringPersonalizationWeb2017}. One way to control for this variation is to hold personalization factors constant while scraping~\parencite[41-42]{metaxaAuditingAlgorithmsUnderstanding2021}. 

A previous study that measured the personalization in Google found that, on average, 11.7\% of its results showed some variation due to personalization~\parencite{hannakMeasuringPersonalizationWeb2017}. Being logged in (or not) impacted the ordering of the results. The machine's IP address led to changes in the results, as that is related to location; results were dramatically different across multiple countries. Significant variations were also found from day to day~\parencite[2-3,12,15,17,19-21]{hannakMeasuringPersonalizationWeb2017}. The analysis also found a \textit{carry-over effect}, a noise from which a previous search may contaminate further queries. An example is that looking for a clothes retailer right after searching "Hawaii" may lead the engine to display stores in that state regardless of where the search occurs. This effect influences queries executed within 10 minutes of each other~\parencite[6-7]{hannakMeasuringPersonalizationWeb2017}. 

\subsection{Data Collection}

The analyzed search topics were selected by extracting the most popular keywords from Google Trends for each country using its official RSS feed. Google Trends lists the most popular search queries, meaning they are relevant to the public. Keywords gathered from Trends were used to query Google Search. This used the \texttt{Selenium} library in Python, which allows for the automation of web browsers. The process used a clean browser every time to avoid contaminating the results with the previous browsing. To simulate a person, the script entered Google's homepage, typed the keyword in the search box using random keystroke times, and then hit "enter," leading to the main results page. The script would then click the "News" tab button and download that page's results. Ten keywords for each country were manually validated and completely matched the results gathered by the system. The keywords that did not include the news tab button in their results page were assumed not to be news-related, verified by manual inspection, and discarded.

The extractions happened every three hours to accommodate the time zones of the countries analyzed while ensuring multiple parts of the news cycle were captured. The data was gathered by simulating users in Brazil, the United Kingdom, and the United States, using \textit{NordVPN}, a leading VPN provider~\parencite{andreastheodorouBestVPNService2024}. All data was collected using an AWS Linux machine. The operating system should not influence search results~\parencite[15]{hannakMeasuringPersonalizationWeb2017}. Whenever the system connected to the VPN, it checked its own IP address using \texttt{ipify.org}'s API. The corresponding location for each IP was later validated using \texttt{ipapi}, which provides an approximated address to an IP. The results for each keyword were also validated to check if they matched the expected language. 

Each keyword was queried five times, thus over 15 hours. This interval was deemed relevant as it gives newsrooms time to react to trending events, even if they happen overnight. Analyzing a longer period would not align with the study's objectives, as it primarily focuses on the news users \textit{receive}. Over time, Google could pick up on additional results for each query, which might suggest better diversity. This could be misleading given that a significant portion of the audience engagement occurs within the first few hours~\parencite{jacknearyWhatLifespanArticle2023}.

Data was collected between March 11\textsuperscript{th}, 2024, and April 1\textsuperscript{st}, 2024. A total of \texttt{2,298} \textit{keywords} were captured, rendering \texttt{143,976} \textit{results}. These account for \texttt{50,496} unique \textit{URLs} from \texttt{4,296} \textit{outlets}. The results included 67 tweets, which were removed. These appeared when searching the names of cities, displaying popular content for that area. 

Using a similar process, additional data was collected for 12 days (April 29\textsuperscript{th} 2024 to May 11\textsuperscript{th} 2024) to validate some of this study's main findings. It includes \texttt{77,887} new \textit{results}. It is described in Appendix~\ref{app:terminology}.

\subsubsection{Carry-over Effect}

In tests for this analysis, the carry-over effect (section~\ref{sec:challenges}) was spotted in Google's main results page as described in previous research~\parencite[6]{hannakMeasuringPersonalizationWeb2017}. This effect should disappear if there is a 10-minute interval between queries~\parencite[7]{hannakMeasuringPersonalizationWeb2017}. However, given the volume of queries in this study, that interval is impractical. Using a computer in Cambridge (UK), the additional tests involved searching for "best pizza places" and examining the results, which were consistent across normal and incognito browsing. After a 10-minute interval, a search for "Train tickets to Manchester" was performed in normal browsing mode, followed by a repetition of the pizza query. This time, a result related to Manchester, which had not appeared before, was shown mid-page in both browsing modes. After another 10-minute interval, the pizza query yielded the original results. A similar methodology was applied to the news tab, where the effect was not observed. Later, the process was repeated using "traffic" as the original query, achieving the same results. Consequently, it can be inferred that the carry-over effect in the news tab, if present, is negligible.

\subsection{Additional Data and Data Labeling}

\subsubsection{News Section}

Using the sample news article provided along with each query on Trends, each keyword was assigned to one of four news sections: Sports, Politics, Entertainment, and Others. This classification system was defined after an initial manual inspection. This process ensures the labeling remains simple, with few categories, while providing the necessary insight for the analysis. The classifications used OpenAI's \texttt{gpt-3.5-turbo-1106} model. The prompt constrained the system to the desired section labels and allowed it to abstain. None was left unassigned. The \textit{temperature} was \texttt{0.0} to increase consistency~\parencite[2]{gilardiChatGPTOutperformsCrowdWorkers2023}.

The labels were validated by manually coding a random sample of 5\% of entries from each country, 125 total. These were annotated by the author and an additional coder, a journalist with experience working in Brazil and the UK. About half the data was classified as Sports, so the inter-coder agreement was tested using Gwet's AC1, a better choice for skewed datasets~\parencite{wongpakaranComparisonCohenKappa2013}. Considering both human coders and GPT, it was \texttt{.898} (95\% Confidence Interval: \texttt{0.824}, \texttt{0.952}), indicating a high level of reliability. A slight drop from when considering only the two human coders, \texttt{0.929} (CI: \texttt{0.876}, \texttt{0.981}).

\subsubsection{Political Bias}\label{sec:detpoliticalbias}

Each outlet in the dataset was classified in regards to its political bias on a seven-degree scale: "far-left," "left," "center-left," center," "center-right," "right," and "far-right." Where available, these ratings were extracted from Media Bias/Fact Check (MBFC). This service specializes in labeling media sources under multiple criteria~\parencite{LeftVsRight2021} and is often cited in academic and media sources~\parencite{akerPredictingNewsSource2019}, \parencite{NoEvidenceDisease2021}, \parencite{resnickIffyQuotientPlatform2018}. There was no information for most sources from Brazil, so these were removed from this part of the analysis. MBFC data accounted for 951 of the 3,082 identified UK and US sources (29.8\%). These were incremented with using GPT-4, following research that showed high correlation between MBFC and the language model in a similar task~\parencite{hernandesLLMsLeftRight2024a}. \texttt{Gpt-4o-2024-05-13} was used to label the remaining sources using the same classification system. The prompt allowed the system to abstain. \textit{Temperature} was set to \texttt{0.0}. These resulted in an additional 307 ratings. A random sample of 10\% of those was validated by the author. These either were Sports outlets categorized as "no political bias" or matched the classification of alternative rating sources, All Sides and Ad Fontes Media. So, in the end, the analysis had \texttt{1,258} (\texttt{40.8\%}) outlets classified in terms of political bias, accounting for \texttt{65,457} (\texttt{72\%}) results.

\subsubsection{Baseline from Media Cloud}\label{sec:mediacloud}

Each keyword in this analysis was queried in the Media Cloud database, a system that monitors and aggregates news output from sources around the world using their RSS feeds, providing a list of texts and sources matching a query~\parencite{robertsMediaCloudMassive2021}, \parencite{FAQs2023}. Media Bias and Google searches, however, do not have an exact correspondence, as Media Cloud looks for string matches within the texts, and Google can index content even if the wording is not the same~\parencite{RankingResultsHow}, \parencite{QueryGuide2023}. Because of that difference, in Media Cloud, keywords of three letters or less were discarded, as these can be too generic. Also, outliers in the volume of results were removed. Each keyword was queried to a Media Cloud collection of national news sources from the relevant country. Data was collected from the day the keyword appeared on Google Trends until the following day to capture a similar news cycle.

\subsection{Attributing Weighted Rankings}\label{sec:methodsweightedrankings}

Rankings in Google results matter. It has been empirically demonstrated that links appearing higher on the page normally have higher click-through rates, are more often seen, and regarded as more trustworthy by users~\parencite{panGoogleWeTrust2007}, \parencite{epsteinSearchEngineManipulation2015}, \parencite{clayLatestClickThroughRate2021}, \parencite{southern25PeopleClick2020}. So, the analysis includes a weighted ranking of the results. To calculate these values, this study borrows the logic from Rank-Biased Overlap (RBO)~\parencite{webberSimilarityMeasureIndefinite2010}. Appearing first yields a weight of 1, and the following entries have progressively decaying scores. The rate of the decay is \( p \) where \( 0 < p < 1 \), and the weight assigned to a URL at position \( k \) is calculated as \( w_k = p^{k-1} \). Here, \( k \) is the rank position, and \( w_k \) is the weight assigned to that position. The value of \( p \) is found by satisfying the equation \( p + p^2 + p^3 = 0.55 \). This way, the top-3 results will account for 55\% of the weights, reflecting the most conservative click-through values recently found by agencies that specialize in optimizing content for Google Search~\parencite{clayLatestClickThroughRate2021}, \parencite{beusWhyAlmostEverything2020}.

\begin{wrapfigure}{r}{0.43\textwidth}
    \centering
    \includegraphics[width=0.41\textwidth]{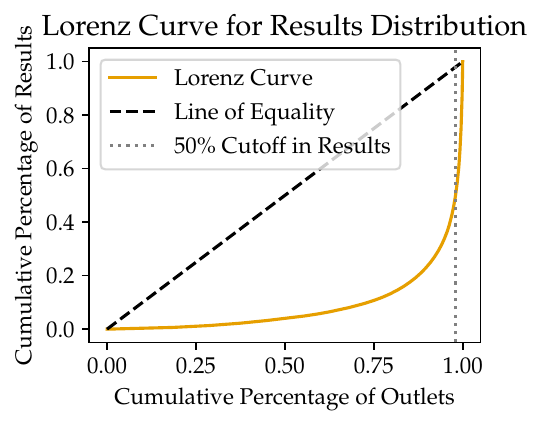}
    \caption{The Lorenz curve shows that a few outlets account for half of the results. The dotted upright line divides the number of results in equal parts.}
    \label{fig:lorenz}
    \vspace{-10pt}
\end{wrapfigure}

\section{Results}\label{sec:results}

\subsection{Source Concentration}\label{sec:sourceconcentration}

The analysis of source concentration in this dataset can be misleading depending on how it is considered. Aggregating the data by keywords, showing the concentration for each query, indicates a balanced environment. These were measured using adapted versions of the Herfindahl-Hirschman Index (HHI), and Gini coefficient. HHI is \texttt{0.09} and Gini is \texttt{0.32}, both on average. These values remain low across the multiple news sections and countries throughout the data.

Analyzing the data horizontally, however, shows that some news outlets get repeated over multiple queries. So, while users might not see repetition in a single search, they might encounter the same sources often over time. In each analyzed country, a small proportion of the outlets account for a large share of the results. These form a non-uniform distribution, as shown by the Kolmogorov-Smirnov statistic (\(D = 0.916\), \(p < 0.001\)). That pattern is also true when looking at the data segmented by section on top of country. A \( \text{CR}_k \) (concentration ratio) analysis shows how concentrated the results are on a small set of outlets, seen in table~\ref{tab:crk}. Despite some fluctuation, these patterns hold true when the data is analyzed by news sections. This breakdown is available in Appendix~\ref{app:numbersnewssection}, as is a list of the most popular outlets per country.

This scenario can also be seen in the overall data having a high Gini coefficient (\texttt{0.822}) and low HHI (\texttt{0.005}). The latter picks up on the fact that several players are represented in the results, but half appear five times or less. So, among these peers that comprise most of the analyzed sources, some leveled competition is going on. Gini, however, picks up on having 2.1\% of these outlets accounting for 50\% of the results, which can be visualized in Figure~\ref{fig:lorenz}.

\begin{table}[htbp]
    \centering
    \caption{\( \text{CR}_k \), concentration of Top \(k\) outlets per country.}\label{tab:crk}
    \begin{tabular}{l|rrrrr}
        \toprule
        \textbf{Country} & \textbf{\( \text{CR}_4 \)} & \textbf{\( \text{CR}_8 \)} & \textbf{\( \text{CR}_{10} \)} & \textbf{Results} & \textbf{News Sources} \\
        \midrule
        BR & 0.163313 & 0.254306 & 0.286720 & 53125 & 1226 \\
        UK & 0.209089 & 0.318496 & 0.357862 & 40619 & 1478 \\
        US & 0.117714 & 0.186734 & 0.214823 & 50232 & 2189 \\
        \bottomrule
    \end{tabular}
\end{table}

\subsubsection{Source Concentration in Privileged Positions}\label{sec:concentrationprivileged}

Considering the ordering of the results, a greater disparity is found. The Gini coefficient is higher (0.919), and HHI remains low (0.011). The concentrated nature of the most privileged positions can also be seen in \( \text{CR}_k \), as per table~\ref{tab:crkweighted}. The source concentration patterns were validated in the additional dataset, which showed similar values for HHI and Gini, and the same pattern of horizontal concentration. It is available in Appendix~\ref{app:additionalcrk}.

\begin{table}[ht]
    \centering
    \caption{\( \text{CR}_k \), concentration of Top \(k\) outlets per country considering the results' weighted ranking.}
    \begin{tabular}{l|rrrrr}
        \toprule
        \textbf{Country} & \textbf{\( \text{CR}_4 \)} & \textbf{\( \text{CR}_8 \)} & \textbf{\( \text{CR}_{10} \)} & \textbf{Results} & \textbf{News Sources} \\
        \midrule
        BR & 0.325873 & 0.651520 & 0.813704 & 53125 & 1226 \\
        UK & 0.325956 & 0.651862 & 0.814496 & 40619 & 1478 \\
        US & 0.306577 & 0.613155 & 0.766245 & 50232 & 2189 \\
        \bottomrule
    \end{tabular}
    \label{tab:crkweighted}
\end{table}

\subsection{Popular Sources are Popular in Google}\label{sec:popularityresults}

Concentration on specific outlets leads to questions about what characteristics they have in common or what leads to news sources appearing in Google more often. Open PageRank scores, understood as a proxy for popularity, showed a clear connection with the results. The average score is higher among the top outlets in the three countries (Table~\ref{tab:uresultsdecimals}). In parts, this is expected since this score is deliberately created to try to mimic Google's Pagerank feature, which is used to determine search results~\parencite{WhatOpenPageRank2023}. 

\begin{table}[htbp]
    \centering
    \caption{Mann-Whitney U test results for Open PageRank scores.}
    \begin{tabular}{lcccc}
        \toprule
        \textbf{Country} & \textbf{Median for Other Outlets} & \textbf{Median for Top Outlets} & \textbf{U-statistic} & \textbf{P-value} \\
        \midrule
        \textbf{Brazil} & 4.08 & 5.27 & 4920.5 & $< 0.001$ \\
        \textbf{UK} & 5.35 & 6.33 & 6872.5 & $< 0.001$ \\
        \textbf{US} & 5.37 & 6.67 & 19634.0 & $< 0.001$\\
        \bottomrule
    \end{tabular}
    \label{tab:uresultsdecimals}
\end{table}

The influence of website popularity on the number of results was analyzed through a set of Ordinary Least Squares (OLS) regressions. Considering only the Politics and Others sections for all countries, a one-point increase in the Open PageRank score is associated with having 14 extra results ($p < 0.001$). Despite the model's limited explanatory power ($R^2 = 0.07$), it provides evidence of this relationship, which can also be seen in the countries individually and considering other news sections. Regression tables are available in Appendix~\ref{app:decimalsregressiontables}. 

\subsection{Information Quality's Impact on Results}\label{ref:infoqualityresults}

In the UK and the US, the top outlets (respectively, the 1.9\% and 3\% with 50\% of the results) have a higher information quality score than other sources. This score was extracted from a dataset created by \textcite{linHighLevelCorrespondence2023a} that rates news sources based on their credibility (referred to here as \textsc{Lin-PC1}). Data for most Brazilian sources was unavailable, so these were removed from this part of the analysis. The information was available for \texttt{1,261} (\texttt{40.9\%}) of the 3,082 UK and US sources. These accounted for \texttt{65,080} (\texttt{71.6\%}) results. To mitigate that gap, the analysis ran on publishers with five results or more (above the median) since those at the bottom often lacked the Lin-PC1 score.

The median Lin-PC1 score for the top outlets was \texttt{0.78} in the UK and \texttt{0.83} in the US, while the rest of the outlets had respectively \texttt{0.70} and \texttt{0.71}. Its statistical relevance was assessed by Mann-Whitney \( U \) (UK: \( U = 3103.5 \), \( p < 0.05 \); US: \( U = 4911.5 \), \( p < 0.01 \)). This comparison only focuses on results in the Politics and Others sections. Including Sports and Entertainment showed a similar pattern but only had statistical significance in the US (\( p < 0.01 \)).

OLS regressions, however, suggest no relationship between the quality rating and an outlet's prevalence when considering popularity. The results indicate that while the Lin-PC1 metric is significantly (negatively) associated with the number of results when controlling for Open PageRank, its direct association with the volume of entries is not significant. This suggests that the apparent association between higher Lin-PC1 and appearing more often in the results may be driven by the relationship between Lin-PC1 and Open PageRank, rather than a direct effect. The regression tables are available in Appendix~\ref{app:linpc1regressiontables}.

\subsection{Political Inclination}\label{sec:polinclination}

Comparing the top outlets in the US to those that had fewer results shows a leftwards inclination, with the distribution's median landing on the center-left instead of the center (Mann-Whitney \( U = 33126.0 \), \( p < 0.05 \)). This difference, however, needs to be taken with a grain of salt, as a significant portion of the data was unclassified. This considers only the Politics and Others sections, which relate more to such bias than Sports and Entertainment. An OLS regression on the scale and the number of results, controlling for popularity (Open PageRank), is not statistically relevant. 

\begin{figure}[ht]
    \centering
    \begin{minipage}[b]{0.45\linewidth}
        \centering
        \includegraphics[width=\linewidth]{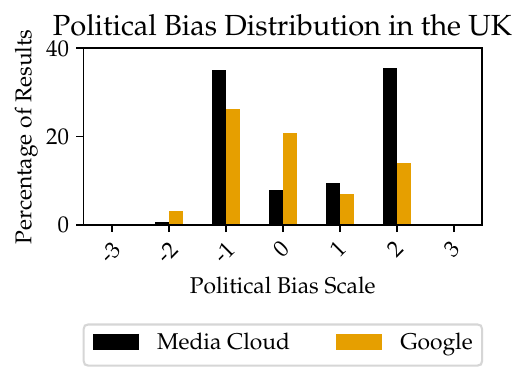}
    \end{minipage}
    \hspace{0.05\linewidth}
    \begin{minipage}[b]{0.45\linewidth}
        \centering
        \includegraphics[width=\linewidth]{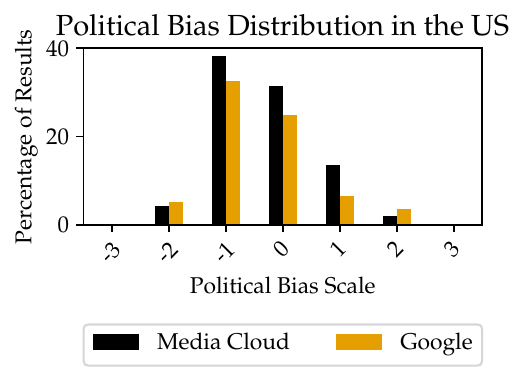}
    \end{minipage}
    \caption{Results per political bias category for Media Cloud's and Google's datasets; -3 is far-left and 3 is far-right. Full table in Appendix~\ref{app:politicalbiasdistribution}.}
    \label{fig:biasscalebycountry}
\end{figure}

More evidence comes from comparing Google Search to Media Cloud's UK and US data (Figure~\ref{fig:biasscalebycountry}). Results in Google gravitate more towards the left categories than Media Cloud's. This pattern relates to a previous study that also identified this left-leaning pattern in the results in the US when comparing to another baseline of news outlets, GDELT~\parencite[9]{trielliSearchNewsCurator2019}. This result backs those found previously and shows this phenomenon in another country.

Moreover, a meaningful relationship exists between political bias and results' rankings. An OLS regression on the ordered political bias (where lower values relate to the left) was run against the order of the results in Google. In the US, it found a significant relationship in which a higher political bias scale rating (more to the right) means appearing further down in Google's results ($ p < 0.001 $). The analysis controlled for popularity. Regression tables are available in Appendix~\ref{app:politicalbiasregressiontables}. Results were not statistically significant in the UK.

The overall leftwards skew can also be verified in individual queries through a time-averaged output bias (TOB) analysis, as described by~\textcite{kulshresthaSearchBiasQuantification2019}. This was applied to the 81 queries in the Politics section in the US and the UK where 80\% or more of results had political bias information. Average \textit{TOB} was -0.05, reinforcing the slight leftwards skew in the results. Most (64) queries were skewed to the left, and they did not necessarily reflect the keywords. For example, queries on left-wing politicians and personalities had both leftwards \textit{TOBs} (Barack Obama, Joe Lieberman, Owen Jones, etc.) and rightwards \textit{TOBs} (Tammy Murphy, Rachel Reeves). The opposite was also true.

\subsection{Preference for Recent Results}\label{sec:recentresults}

Unsurprisingly, the analysis reveals a pattern of updating the content to incorporate more recent results over time. That can be seen in the rate at which average article age progresses (Table~\ref{tab:analysisoveriterations}). Given that this study gathered data for each query every three hours, with no changes, the average from a second extraction should be three hours older than the first. These being smaller than three hours reveals that the results prioritize newer articles. Recency analysis removed outliers in terms of article age using the IQR method. The same pattern was found in every country.

Jaccard index and Rank-Biased Overlap (RBO) scores are lower when comparing the first and last iterations to subsequent ones (1 vs. 2, 2 vs. 3,\dots), suggesting a gradual change. The results between the initial and final scraping sessions (1 and 5) reveal a medium level of overlap (Table~\ref{tab:jaccardanalysis}). 

\begin{table}[ht]
    \centering
    \caption{Descriptive statistics of Jaccard Index and RBO comparing first iteration to last.}
    \begin{tabular}{lccccc}
    \toprule
    \textbf{Statistic} & \textbf{Count} & \textbf{Mean} & \textbf{Std. Deviation} & \textbf{Minimum} & \textbf{Maximum} \\
    \midrule
    Jaccard & 2298 & 0.491 & 0.198 & 0.000 & 1.000 \\
    RBO & 2298 & 0.347 & 0.379 & 0.000 & 0.999 \\
    \bottomrule
    \end{tabular}
    \label{tab:jaccardanalysis}
\end{table}

A steady pattern of content renewal is true for all analyzed countries (Table ~\ref{tab:analysisoveriterations}). The average sequential Jaccard on the outlets' names shows a higher value (0.65) than considering the URLs (0.54), suggesting that part of the variability does not bring additional source diversity. An ANOVA test shows no meaningful difference in Open PageRank over iterations. So, despite changes in the URLs, there is no difference in the popularity of the featured outlets over time.

\begin{table}[ht]
\centering
\caption{Average Jaccard, RBO, and time difference between article publishing and data collection (Article Age) over different iterations. Jaccard and RBO scores represent a comparison with the iteration that came immediately before. The mean variations in the columns were tested using ANOVA and are all statistically significant (\(p < 0.001\)).}
\begin{tabular}{lccc}
\toprule
\textbf{Scraping Iteration} & \textbf{Avg. Jaccard} & \textbf{Avg. RBO} & \textbf{Avg. Article Age (Hours)} \\
\midrule
1 &      &      & 14.09 \\
2 & 0.51 & 0.53 & 14.70 \\
3 & 0.53 & 0.55 & 15.76 \\
4 & 0.56 & 0.58 & 16.86 \\
5 & 0.56 & 0.58 & 17.68 \\
\bottomrule
\end{tabular}
\label{tab:analysisoveriterations}
\end{table}

The analysis reinforces the importance of recency for the articles. Outlets were divided into four groups based on their popularity (Open PageRank) and how they performed regarding their number of results, as seen in Table~\ref{tab:recencycategories}. Outlets in the top half of the results volume had a lower average article age than their counterparts in the bottom half. This shows that less-known sources could get extra space by having newer content. 

The source concentration patterns were validated in the additional dataset, which showed similar values to the ones displayed here. It is available in Appendix~\ref{app:validationdataset}.

\begin{table}[ht]
    \centering
    \caption{Average and median article age for different categories of outlets regarding their popularity and volume of results. Mean variations tested using ANOVA (\(p < 0.001\)) (\(p < 0.001\))}
    \begin{tabular}{llrr}
        \toprule
        \textbf{Country} & \textbf{Category} & \textbf{Avg. Article Age} & \textbf{Median Article Age} \\
                        &                   & (Hours) & (Hours) \\
        \midrule
        \multirow{4}{*}{\textbf{BR}} & Low Popularity, Low Results & 17.2 & 15.0 \\
        & Low Popularity, High Results & 15.3 & 14.0 \\
        & High Popularity, Low Results & 16.7 & 15.0 \\
        & High Popularity, High Results & 15.8 & 14.0 \\
        \hline
        \multirow{4}{*}{\textbf{UK}} & Low Popularity, Low Results & 17.2 & 16.0 \\
        & Low Popularity, High Results & 15.1 & 13.0 \\
        & High Popularity, Low Results & 17.4 & 17.0 \\
        & High Popularity, High Results & 14.5 & 12.0 \\
        \hline
        \multirow{4}{*}{\textbf{US}} & Low Popularity, Low Results & 16.4 & 14.0 \\
        & Low Popularity, High Results & 15.9 & 15.0 \\
        & High Popularity, Low Results & 15.2 & 13.0 \\
        & High Popularity, High Results & 13.8 & 11.0 \\
        \bottomrule
    \end{tabular}
    \label{tab:recencycategories}
\end{table}

On average, articles ranking higher are newer. This relationship can be seen in an OLS regression on the weighted order of the articles and article age, controlling for the article's popularity ($R^2 = 0.024, p < 0.001$). This is true for every country. Their tables are in Appendix~\ref{app:recencyregressiontables}.

\begin{table}[ht]
    \centering
    \caption{Article recency for categories of positions in Google's ranking. Mean variation tested using ANOVA (\(p < 0.001\)).}
    \begin{tabular}{lccc}
    \toprule
    \textbf{Position} & \textbf{Number of texts} & \textbf{Avg. Article Age (Hours)} & \textbf{Median Article Age (Hours)} \\
    \midrule
    \textbf{1}   & 11250 & 13.3 & 12.0 \\
    \textbf{2-5} & 43651 & 14.0 & 12.0 \\
    \textbf{6+}  & 77205 & 17.2 & 15.0 \\
    \bottomrule
    \end{tabular}
    \label{tab:orderandtime}
\end{table}

\subsection{Publishing More About a Topic}\label{sec:publishingmore}

Given Google's preference for recent content, news outlets might be incentivized to publish multiple texts about a topic to appear more often in the results. Comparing Google's results to Media Cloud's suggests this might work. Outlets that had more texts on a given query in Media Cloud had increased chances of being listed in Google's results. This was confirmed by logistic regressions (\(R^2_{McF} = 0.04, p < 0.001\)). Overall, each additional text on a keyword that appeared on Media Cloud's database meant a 10.4\% greater chance of that outlet appearing in Google's results for that query. Tables are available in Appendix~\ref{app:mediacloudregressiontables}.

\section{Discussion}\label{sec:discussion}

Google's search algorithm's complexity makes it hard to discern patterns in its selection and ranking of news content. Its engine is speculated to rely on several attributes, which are unknown to the public~\parencite{kingSecretsAlgorithmGoogle2024}, \parencite{satoBiggestFindingsGoogle2024}. This opacity makes approaches like the algorithm audits necessary for capturing some of the issues in its workings by controlling the inputs and analyzing its outputs~\parencite[197]{kulshresthaSearchBiasQuantification2019}. Given said complexity, some of the discrepancies found in this investigation hinge on a slight difference or models showing a low R-squared, which, despite statistically meaningful, only account for a small portion of the variation in the data. Though, with Google's relevance, the tiniest nudge can have a big impact. 

The results displayed in Google's news tab are mostly balanced vertically but concentrated horizontally. In other words, a user who looks for a news topic in Google may see no repeated outlets (this happened in 40\% of the queries in this audit). The Gini coefficient was low when considering individual queries, indicating low market concentration (section~\ref{sec:sourceconcentration}). However, users who are too dependent on Google for their news diets, as people increasingly are~\parencite{newmanReutersInstituteDigital2023}, will often see repeated outlets across their queries. This can be seen in a large Gini overall and relates to research that found these platforms bring more short-term consumption of diverse news sources but limit it in the long term~\parencite{jurgensMappingExposureDiversity2022}.

It should be noted that the low Gini for individual queries is not always true. Some cases of high concentration happen when looking for outlets' names. This could be justified as the users' interest to see mostly \textit{CNN} when searching for "CNN." However, others refer to queries on topics such as an Uber driver strike in Brazil, for which five outlets are featured among 50 results (over five iterations). 

Google's source concentration had already been demonstrated in another analysis of Google's news results~\parencite{trielliSearchNewsCurator2019}. This analysis shows it goes beyond the US, as the same pattern was identified in the UK, similar to the US regarding language and development, and Brazil, with a different language and situated in the Global South. This shows the global nature of Google's impact, a power that extends beyond any publisher's dream~\parencite[173]{nielsenPlatformPower2022}.

In practice, search engines act as gatekeepers to the content that gets consumed or not on the web, to the point that not being listed in Google can be seen as the same as not existing online~\parencite[21]{hinmanEsseEstIndicato2005}. By narrowing the options users see, these results might affect the internet's value as a source of information, which comes from being a democratic and inclusive place that is able to give voice to diverse groups~\parencite[169]{lucasd.intronaShapingWebWhy2000}, \parencite[180-181]{lucasd.intronaShapingWebWhy2000}, \parencite[11]{diazGoogleGogglesSociopolitical2008}. The top outlets are already popular (section~\ref{sec:biasdistortions}). The cost of biased search mechanisms can be significant for those striving to be found.

Sources that offer quality content but lack popularity might end up neglected. Someone seeking information about a news event may stop their search after finding a suitable story near the top of the list, missing equally valuable options listed further~\parencite[180]{lucasd.intronaShapingWebWhy2000}. This concentration might go unnoticed by the user; after all, their information needs were met. However, it might still have the consequence of leaving them systematically consuming from similar sources (relating to the filter-bubble effect~\parencite{haimBurstFilterBubble2018}) while leaving many potentially quality sources unseen.

\subsection{Popularity Bias and Distortions}\label{sec:biasdistortions}

The concentration privileges certain characteristics. One of these is popularity, measured by the Open PageRank score, with Google favoring well-known outlets. This approach has its merits, as it ends up giving more space to reputable sources (section~\ref{ref:infoqualityresults}). Previous research indicates that the search engine implicitly considers credibility through popularity and honors it to a point~\parencite[3,10]{lewandowskiCredibilityWebSearch2012}. This can be seen empirically in this analysis, as the outlets featured the most also scored higher on Lin-PC1. In consonance with previous work, however, this study shows that the higher-quality domain prevalence is incidental.

Popularity does not necessarily equate to quality, though. Sometimes, widely read sources may disseminate lower-quality information. For instance, English tabloids, which are highly popular, score lower on information quality metrics. The tabloid \textit{Daily Mail} is, in the UK, the third most consumed online publication and the largest paid newspaper in terms of circulation among those that publicize the metric~\parencite[]{tobittNationalPressABCs2024}, \parencite[59]{newmanReutersInstituteDigital2023}. However, it has a low score in the Lin-PC1 metric (0.38). In this analysis, it was the second most popular source in the UK regarding total results count and third when weighting in the ordering of the articles. It featured in the top 10 outlets for three of the four news sections: Others, Sports, and Entertainment. It appeared more often than sources with higher Lin-PC1, such as \textit{The Guardian} (0.75), \textit{The Independent} (0.73), \textit{Telegraph} (0.61), and \textit{Sky News} (0.87). The \textit{Daily Mail}'s case is not isolated, with other low Lin-PC1 sources featuring in the UK's top-20 results. \textit{Daily Mail} also ranks high in the US.

And then, are these sources not part of the news ecosystem and, thus, deserving of inclusion in these results? This is a highly contentious matter. Taking a marketplace of ideas approach, one might argue that all sources, regardless of their quality score, should be included in the debate~\parencite[236]{gordonJohnStuartMill1997}. An algorithm that rewards popularity promotes equality by using consumption to make decisions. By basing its rankings on popularity, Google may attempt to exempt itself from making explicit editorial decisions about which outlets deserve more space. After all, public demand drives the rankings. The presence of a wide range of sources would encourage competition and eventually lead to a selection of higher-quality information~\parencite{gordonJohnStuartMill1997}. However, this approach has been criticized for various reasons. One of them is suffering from failures such as monopolies, as in economic markets, cornering minority views~\parencite[]{goldmanSpeechRegulationMarketplace1999}, \parencite[237-241]{gordonJohnStuartMill1997}, \parencite[109, 112]{herzogWhatWrongMarketplace2023}, \parencite[]{helbergerDemocraticRoleNews2019} ---which, here, can be seen as the dominance of sources that are already popular.

The imbalance can be seen in the prioritization of national, legacy outlets. This preference also means being prone to reproduce their biases. If traditional media is skewed to the left, the algorithm will reproduce this bias. Indeed, this exact pattern was found. In the US, the content output of outlets classified as national by Media Cloud was slightly skewed to the left. A similar pattern was found in Google. Meanwhile, in the UK, Media Cloud's content was more balanced, which was reflected in Google's data, although with some left skew. This investigation failed to find a meaningful direct relationship between political bias and featuring more in the results. So, this behavior could originate from the system reacting to the audience consuming more content aligned to the left or from this political bias being related to other algorithm parameters. Despite what caused the leftwards skew, considering news diversity in terms of the content "as received" by the audience, the aspect that matters most here is the distribution itself ---and these results highlight a preference.

That comes on top of a seemingly systematic leftwards bias. It is present for most of the analyzed politics-related queries, independently of biases in the entity represented in the keyword, and can impact results' rankings (section~\ref{sec:polinclination}). This contrasts with previous research that considered the main results page of Google and found that it matched the analyzed entity (left politicians had a left skew, right had a right skew). In that case, the research pointed to candidate-controlled sources, such as their own websites, pulling the results~\parencite[215-217]{kulshresthaSearchBiasQuantification2019}. Removing those to see only news content makes the preference in the algorithm's output more apparent.

Also, Google itself has a significant influence on news sources' traffic~\parencite[10]{newmanReutersInstituteDigital2023}, \parencite[]{GlobalAudienceInsights2024}. A popularity-based approach can, thus, exacerbate existing inequalities and lead to a greater divide between more and less privileged outlets, which can be problematic~\parencite[180]{lucasd.intronaShapingWebWhy2000}. By promoting certain sources, Google can create a feedback loop where larger, more popular brands increasingly get more space and visibility, further entrenching their dominance. This creates another distortion in the media landscape by amplifying certain publishers, making it harder for smaller ones to gain traction. In practice, there is already a noted bias towards national over local media, which was also found in previous research~\parencite[13]{trielliSearchNewsCurator2019}, and in local publisher's perception of the matter, who feel like an "afterthought" for tech platforms~\parencite[162]{nielsenPlatformPower2022}. 

This speaks to an ongoing problem in the media landscape: the decline of local journalism. The lack of diversity in the sources promoted by Google's algorithm can have nefarious effects on the outlets. Financially, it can limit the revenue opportunities for smaller, local, or less popular outlets, potentially leading to their decline or closure. In the past, changes in Facebook's feed ranking led to drops in traffic to news outlets, bringing layoffs and even causing a digital publisher to shut down~\parencite[]{mikeshieldsFacebookAlgorithmHas2018}, \parencite[]{elliottFacebookMadeBuzzFeed2023}, \parencite[185]{nielsenPlatformPower2022}. This raises concerns about the relationship between source diversity and larger issues in the media landscape, such as the local news deserts, areas that lack news outlets to cover issues in that community. These smaller outlets face economic challenges caused by the shift to online news, and their absence hampers areas like the efficiency of local governments, citizen engagement, and participation in elections~\parencite[5]{napoliAssessingLocalJournalism2018}. In Brazil, for example, 13.2\% municipalities lack local journalism~\parencite{sergioludtkeBrasilTemReducao2023}.

Google's algorithm shows some distortions that must be considered. Being a reputable and national source does not guarantee that the search will be successful, and Brazil's results can serve as an example. The list of top-ranking sources excludes some outlets of high Open PageRank scores and national relevance, such as \textit{Veja}. It is the most important weekly magazine in the country and displays conservative values~\parencite{marcioemiliodossantosAnalysingBigCultural2012}, \parencite{fernandaargoloWomanOutPlace2018}, \parencite{hannayahyaRevistasEm20212022}. It was featured in 27 queries, while Media Cloud data shows that it produced content on at least 19 additional topics. Also, \textit{Correio Braziliense}, the 9\textsuperscript{th} newspaper in Brazil in terms of circulation, outnumbers both \textit{O Estado de S. Paulo} (3\textsuperscript{rd}) and \textit{Folha de S.Paulo} (1\textsuperscript{st}), while having, respectively, 8\% and 2\% of their number of subscribers~\parencite{ComAssinaturaBarata2024}.

Outlets might game the algorithm to secure better spots through practices such as Search Engine Optimization, a combination of technical and content manipulation that boosts Google's algorithm rating for a page~\parencite[5]{lewandowskiCredibilityWebSearch2012}, \parencite[165]{nielsenPlatformPower2022}. It can help the discovery of helpful content, but may also negatively impact the credibility of the results~\parencite[5-6]{lewandowskiCredibilityWebSearch2012}. After all, high-ranking results might surface not because they come from the most reputable source but because they are published in a way that scores points with the sorting mechanism~\parencite[175]{lucasd.intronaShapingWebWhy2000}.

The \textit{BBC} provides an interesting case for analysis. It is the only outlet to feature among the top sources in Brazil, the UK, and the US. The \textit{BBC} is the leading news outlet in the UK and among the top in multiple countries~\parencite{newmanReutersInstituteDigital2023}. 

However, the \textit{BBC}'s popularity measured by Google might have been artificially boosted. Its presence in Brazil expanded in the early 2000s with the creation of a website in Portuguese. Today, between 60\% and 65\% of \textit{BBC Brasil}'s traffic comes from search~\parencite[83]{gomesBBCNewsBrasil2021}. The company increased its reach through partnerships that allow major (competing) Brazilian outlets to republish its content like that of a news agency~\parencite[16,155-167]{gomesBBCNewsBrasil2021}. The articles include a \textit{canonical} hyperlink pointing to the original URL in the \textit{BBC}'s website, which can be seen in the republished articles pages' source codes. It tells Google's system that the URLs are duplicates, redirecting its readings of engagement to the \textit{BBC}'s original content, making it seem more popular~\parencite{WhatURLCanonicalization2024}. This analysis does not intend to establish a causal relationship between this practice and being featured often in Google, only to offer it as an example of a practical way an outlet might try to inflate its algorithmic relevance.

\subsection{Recency}

Recency is another significant factor in Google's results. They show a steady rate of churn in the results list while more recent articles are added (section~\ref{sec:recentresults}). Valuing recency might make sense given the fast-paced nature of digital age communications~\parencite[129-130]{OECDReportPublic2021}. However, this approach is not without its drawbacks. It can stimulate the spamming of articles to regularly have something new for the algorithm to pick up (section~\ref{sec:publishingmore}).

Valuing recency makes sense for breaking news but not so much for other topics. The data shows that both breaking news and less urgent news topics appear among the ones with the lowest average article age rates. For the first, an example is the coverage of the Francis Scott Key Bridge collapse in Baltimore (US) from a collision with a ship~\parencite{holmesBaltimoreBridgeCollapse2024}. Its unexpectedness contrasts with St. Patrick's Day, also among the lowest average recency. Balancing shelf lives is a known difficulty in algorithms that distribute news content, as measuring the evergreen-ness of an article, and criteria such as the depth and density it conveys is hard~\parencite[198]{diakopoulosAutomatingNewsHow2019}.

Once again, algorithmic preferences might play a role in media's behavior. In this case, it may affect less popular outlets disproportionately, as these appear less often in the results, and the data shows recency is crucial to break through the algorithm. This pattern might offer a fighting chance for smaller outlets, but it can also force a favoring of breaking news over in-depth analytical reporting or publishing irrelevant or inaccurate updates to have recent content for the search engine to pick up~\parencite{carmenarroyonietoGoogleNewsChanges2019}. 

\section{Conclusion}\label{sec:conclusion}

This study sheds light on the critical influence of Google's search algorithm on news diversity across different cultural and linguistic contexts. It expands previous studies in scale and by analyzing a non-English-speaking country. It also confirms its main findings with an additional dataset, providing encouraging evidence of the generalizability of this study's results. By revealing patterns of source concentration, popularity bias, and slight political skewness, the research underscores the algorithm's role in shaping public access to information. These findings have implications for media diversity, democratic engagement, and the potential perpetuation of existing inequalities within the news ecosystem. Moving forward, there is an urgent need for greater transparency in algorithmic processes and for policy interventions that promote a more equitable distribution of news sources. Given the search engine's characteristic of driving traffic and financial gains, the divide between outlets favored by the algorithm and those neglected might increase. 

While it might seem intuitive that popular outlets receive more visibility, the extent of this concentration---and its consistency across different countries and news sections---is not trivial. The empirical evidence demonstrates that a small fraction of outlets dominate the search results in multiple countries. Moreover, the preference for recent content over potentially more in-depth or analytical pieces raises questions about the quality of information being prioritized.

\subsection{Limitations and Future Work}

The queries used in this study often refer to broader subjects, and about half are related to Sports. Part of the skewness towards national outlets found in this analysis may come from inquiring about topics of these natures. Further studies could benefit from exploring different topics and experimenting with their phrasing, which might impact results~\parencite[190]{kulshresthaSearchBiasQuantification2019}. This study speculates that, as a popularity provider, Google itself might increase the divide between major and less-known outlets over time. Long-term evidence is necessary to understand if that is the case. 

\subsection*{Acknowledgements}

The contributions of Giovanni Bello in data labeling are gratefully acknowledged.

\subsection*{Competing Interests}

The authors declare no competing interests.

\subsection*{Funding}

No funding was received for conducting this study.

\subsection*{Ethics Approval}

This project's data collection was approved by the Leverhulme Centre for the Future of Intelligence Research Ethics Committee.

\subsection*{Data availability}

The datasets generated during and/or analysed during the current study are not publicly available due to being extracted from a private, third-party, source, as established by its ethics approval.

\subsection*{Author contributions}

RH designed the study, RH performed the analyses, and both authors assisted in the revision of the manuscript and the refinement of its arguments.

\printbibliography

\newpage
\begin{appendices}
\section*{Appendix}
\section{Terminology}\label{app:terminology}

The object of this analysis is a dedicated page for news content displayed in Google's search engine, or \textsc{Google Search}. When looking for something on Google, users can access it by clicking "News" below the search box on the results page. That leads to another results page containing only news items. This analysis is concerned with the results from this page, here referred to as the \textsc{news tab}. This is not to be confused with \textsc{Google News}, a news aggregator (a platform that collects news from various sources) and a separate product with search functionalities.

Searching for something in Google's engine may be called making a \textsc{query}, and the words or phrases that are looked up are \textsc{keywords}. The keywords used when querying Google Search come from yet another product, \textsc{Google Trends}, an official service that lists what are the most popular search queries in a country on a particular date. Every news article that appear in these count as a \textsc{result}.

\newpage
\section{Number of Outlets, Queries, and Results per news section}\label{app:numbersnewssection}

\begin{table}[!ht]
    \centering
    \caption{Number of outlets, queries, and URLs for each news section in each country.\textit{Total URLS (Once per query)} ignores repetition of results that appear more than once for the same query (as they were executed multiple times), but will count it again if it also appears as a result for a different query.}
    \begin{tabular}{llrrrr}
    \toprule
     &  & \textbf{Unique Outlets} & \textbf{Unique Queries} & \textbf{Unique URLs} & \textbf{Total URLs} \\
    \textbf{Country} & \textbf{News Section} &  &  &  & (\textit{once per query}) \\
    \midrule
    \multirow[t]{4}{*}{\textbf{BR}} & \textbf{Total} & \textbf{1226} & \textbf{866} & \textbf{18157} & \textbf{21058}  \\
    & Entertainment & 479 & 186 & 3588 & 3950 \\
    & Others & 768 & 180 & 3763 & 4292 \\
    & Politics & 260 & 51 & 1417 & 1466 \\
    & Sports & 579 & 449 & 9659 & 11350 \\
    \midrule
    \multirow[t]{4}{*}{\textbf{UK}} & \textbf{Total} & \textbf{1478} & \textbf{662} & \textbf{14150} & \textbf{16320} \\
    & Entertainment & 624 & 213 & 4075 & 4415 \\
    & Others & 707 & 132 & 2738 & 3142 \\
    & Politics & 236 & 47 & 1108 & 1146 \\
    & Sports & 603 & 270 & 6532 & 7617 \\
    \midrule
    \multirow[t]{4}{*}{\textbf{US}} & \textbf{Total} & \textbf{2189} & \textbf{770} & \textbf{19508} & \textbf{22906} \\
    & Entertainment & 560 & 113 & 2725 & 3004 \\
    & Others & 1020 & 158 & 3657 & 4284 \\
    & Politics & 320 & 46 & 1224 & 1317 \\
    & Sports & 1306 & 453 & 12134 & 14301 \\
    \midrule
    \bottomrule
    \end{tabular}
\end{table}

\newpage
\section{Most Popular Outlets per Country and News Section}\label{app:topoutlets}

\begin{longtable}{lllr}
    \caption{Top 10 Outlets per Country and News Section} \\
    \textbf{Country} & \textbf{News Section} & \textbf{Outlet} & \textbf{Count} \\
    \hline
    \endfirsthead

    \multicolumn{4}{c}{{\tablename\ \thetable{} -- continued from previous page}} \\
    \textbf{Country} & \textbf{News Section} & \textbf{Outlet} & \textbf{Count} \\
    \hline
    \endhead

    \hline \multicolumn{4}{r}{{Continued on next page}} \\
    \endfoot

    \hline
    \endlastfoot

    \textbf{BR} & \textbf{Sports} & ge.globo.com & 2855 \\
    &  & terra.com.br & 1342 \\
    &  & lance.com.br & 1331 \\
    &  & gazetaesportiva.com & 1222 \\
    &  & espn.com.br & 878 \\
    &  & uol.com.br & 808 \\
    &  & cnnbrasil.com.br & 732 \\
    &  & trivela.com.br & 635 \\
    &  & itatiaia.com.br & 582 \\
    &  & noataque.com.br & 570 \\
    & \textbf{Others} & g1.globo.com & 571 \\
    &  & terra.com.br & 358 \\
    &  & cnnbrasil.com.br & 287 \\
    &  & correiobraziliense.com.br & 255 \\
    &  & oglobo.globo.com & 213 \\
    &  & noticias.uol.com.br & 200 \\
    &  & diariodonordeste.verdesmares.com.br & 196 \\
    &  & www1.folha.uol.com.br & 192 \\
    &  & infomoney.com.br & 177 \\
    &  & exame.com & 176 \\
    & \textbf{Entertainment} & terra.com.br & 712 \\
    &  & uol.com.br & 490 \\
    &  & cnnbrasil.com.br & 479 \\
    &  & g1.globo.com & 415 \\
    &  & metropoles.com & 411 \\
    &  & gshow.globo.com & 358 \\
    &  & odia.ig.com.br & 271 \\
    &  & estadao.com.br & 230 \\
    &  & revistaquem.globo.com & 230 \\
    &  & diariodonordeste.verdesmares.com.br & 206 \\
    & \textbf{Politics} & cnnbrasil.com.br & 234 \\
    &  & g1.globo.com & 157 \\
    &  & oglobo.globo.com & 131 \\
    &  & terra.com.br & 128 \\
    &  & poder360.com.br & 126 \\
    &  & metropoles.com & 115 \\
    &  & www1.folha.uol.com.br & 108 \\
    &  & gauchazh.clicrbs.com.br & 93 \\
    &  & estadao.com.br & 91 \\
    &  & oantagonista.com.br & 87 \\
   \hline
   \textbf{UK} & \textbf{Sports} & bbc.co.uk & 1372 \\
    &  & skysports.com & 1000 \\
    &  & theguardian.com & 850 \\
    &  & dailymail.co.uk & 842 \\
    &  & mirror.co.uk & 534 \\
    &  & independent.co.uk & 490 \\
    &  & thesun.co.uk & 461 \\
    &  & telegraph.co.uk & 455 \\
    &  & bbc.com & 378 \\
    &  & standard.co.uk & 349 \\
    & \textbf{Others} & bbc.co.uk & 483 \\
    &  & theguardian.com & 353 \\
    &  & dailymail.co.uk & 348 \\
    &  & independent.co.uk & 335 \\
    &  & news.sky.com & 245 \\
    &  & telegraph.co.uk & 194 \\
    &  & thesun.co.uk & 152 \\
    &  & mirror.co.uk & 135 \\
    &  & cnn.com & 135 \\
    &  & bbc.com & 132 \\
    & \textbf{Entertainment} & dailymail.co.uk & 1064 \\
    &  & independent.co.uk & 597 \\
    &  & mirror.co.uk & 523 \\
    &  & bbc.co.uk & 484 \\
    &  & theguardian.com & 471 \\
    &  & thesun.co.uk & 445 \\
    &  & news.sky.com & 272 \\
    &  & express.co.uk & 251 \\
    &  & telegraph.co.uk & 249 \\
    &  & hollywoodreporter.com & 204 \\
    & \textbf{Politics} & theguardian.com & 290 \\
    &  & bbc.co.uk & 271 \\
    &  & news.sky.com & 182 \\
    &  & independent.co.uk & 172 \\
    &  & telegraph.co.uk & 163 \\
    &  & bbc.com & 86 \\
    &  & thetimes.co.uk & 74 \\
    &  & dailymail.co.uk & 71 \\
    &  & ft.com & 69 \\
    &  & express.co.uk & 62 \\
   \hline
   \textbf{US} & \textbf{Sports} & espn.com & 1624 \\
    &  & cbssports.com & 1603 \\
    &  & si.com & 1219 \\
    &  & sports.yahoo.com & 892 \\
    &  & theathletic.com & 790 \\
    &  & usatoday.com & 658 \\
    &  & nypost.com & 611 \\
    &  & bleacherreport.com & 517 \\
    &  & ncaa.com & 507 \\
    &  & foxnews.com & 398 \\
    & \textbf{Others} & usatoday.com & 410 \\
    &  & cbsnews.com & 305 \\
    &  & cnn.com & 261 \\
    &  & nytimes.com & 246 \\
    &  & apnews.com & 184 \\
    &  & nbcnews.com & 174 \\
    &  & nypost.com & 142 \\
    &  & bbc.com & 137 \\
    &  & washingtonpost.com & 136 \\
    &  & foxnews.com & 128 \\
    & \textbf{Entertainment} & people.com & 483 \\
    &  & yahoo.com & 240 \\
    &  & usatoday.com & 192 \\
    &  & variety.com & 181 \\
    &  & hollywoodreporter.com & 161 \\
    &  & nytimes.com & 148 \\
    &  & dailymail.co.uk & 141 \\
    &  & usmagazine.com & 139 \\
    &  & foxnews.com & 136 \\
    &  & nypost.com & 127 \\
    & \textbf{Politics} & cnn.com & 169 \\
    &  & foxnews.com & 146 \\
    &  & nytimes.com & 136 \\
    &  & nbcnews.com & 120 \\
    &  & thehill.com & 99 \\
    &  & washingtonpost.com & 96 \\
    &  & apnews.com & 92 \\
    &  & politico.com & 90 \\
    &  & msnbc.com & 82 \\
    &  & usatoday.com & 81 \\
\end{longtable}

\newpage
\section{Concentration Ratio per country and news section}\label{app:additionalcrk}

\begin{table}[ht]
    \centering
    \caption{\( \text{CR}_k \) shows the concentration of Top \(k\) outlets for each news section in Brazil, the UK, and the US.}
    \begin{tabular}{llrrrrr}
        \toprule
        \textbf{Country} & \textbf{News Section} & \textbf{\( \text{CR}_4 \)} & \textbf{\( \text{CR}_8 \)} & \textbf{\( \text{CR}_{10} \)} & \textbf{Total Count} & \textbf{Num News Sources} \\
        \midrule
        \textbf{Brazil} & Entertainment & 0.197848 & 0.317727 & 0.358882 & 10594 & 479 \\
        & Others & 0.133545 & 0.206264 & 0.238311 & 11015 & 768 \\
        & Politics & 0.191854 & 0.322314 & 0.374852 & 3388 & 260 \\
        & Sports & 0.239974 & 0.348514 & 0.389470 & 28128 & 579 \\
        \midrule
        \textbf{UK} & Entertainment & 0.219534 & 0.337941 & 0.375216 & 12153 & 624 \\
        & Others & 0.192011 & 0.283782 & 0.317533 & 7911 & 707 \\
        & Politics & 0.317488 & 0.454198 & 0.499653 & 2882 & 236 \\
        & Sports & 0.229955 & 0.339727 & 0.380863 & 17673 & 603 \\
        \midrule
        \textbf{US} & Entertainment & 0.155549 & 0.239143 & 0.276469 & 7046 & 560 \\
        & Others & 0.122200 & 0.185900 & 0.212300 & 10000 & 1020 \\
        & Politics & 0.193559 & 0.321356 & 0.376610 & 2950 & 320 \\
        & Sports & 0.176545 & 0.261741 & 0.291672 & 30236 & 1306 \\
        \bottomrule
    \end{tabular}
\end{table}

\newpage
\section{Distribution of Outlets and Political Bias}\label{app:politicalbiasdistribution}

\begin{table}[!ht]
    \centering
    \caption{Number of outlets per political classification per country. They go from -3 to +3, or far-left to far-right: far-left, left, center-left, center, center-right, right, far-right. There is a column for unclassified sources. Brazil was excluded as it did not have enough outlets tagged. }
    \begin{tabular}{llrrrrrrrr}
        \toprule
         &  & \textbf{-3} & \textbf{-2} & \textbf{-1} & \textbf{0} & \textbf{1} & \textbf{2} & \textbf{3} & \textbf{Uncl.} \\
        \textbf{Country} & \textbf{News Section} &  &  &  &  &  &  &  &  \\
        \midrule
        \multirow[t]{4}{*}{\textbf{BR}} & Entertainment & 1 & 33 & 21 & 12 & 56 & 6 & 0 & 3821 \\
        & Others & 0 & 42 & 41 & 57 & 85 & 13 & 0 & 4054 \\
        & Politics & 1 & 68 & 27 & 35 & 71 & 6 & 0 & 1258 \\
        & Sports & 1 & 21 & 151 & 1486 & 39 & 16 & 0 & 9636 \\
        \midrule
       \multirow[t]{4}{*}{\textbf{UK}} & Entertainment & 0 & 272 & 1450 & 588 & 306 & 774 & 0 & 1025 \\
        & Others & 0 & 118 & 933 & 692 & 272 & 355 & 0 & 772 \\
        & Politics & 1 & 53 & 430 & 270 & 128 & 159 & 0 & 105 \\
        & Sports & 0 & 96 & 1384 & 1773 & 407 & 950 & 0 & 3007 \\
        \midrule
       \multirow[t]{4}{*}{\textbf{US}} & Entertainment & 0 & 445 & 1118 & 505 & 192 & 151 & 0 & 593 \\
        & Others & 0 & 203 & 1599 & 1271 & 438 & 153 & 0 & 620 \\
        & Politics & 0 & 182 & 617 & 236 & 130 & 92 & 1 & 59 \\
        & Sports & 1 & 206 & 3950 & 3983 & 714 & 411 & 0 & 5036 \\
        \midrule
       \bottomrule
       \end{tabular}
\end{table}

\begin{table}[!ht]
    \centering
    \caption{Amount of results that fall into each political bias category for both Media Cloud's and Google's results. These go from "far-left" (-3) to far-right (3).}
    \begin{tabular}{@{}lcccccccc@{}}
    \toprule
     & \multicolumn{8}{c}{\textbf{Bias Scale}} \\ \cmidrule(l){2-9} 
    \textbf{Category} & \textbf{-3} & \textbf{-2} & \textbf{-1} & \textbf{0} & \textbf{1} & \textbf{2} & \textbf{3} & \textbf{Unassigned} \\ \midrule
    \textbf{UK} & & & & & & & & \\
    & & & & & & & & \\
    \textbf{M. Cloud}  & 9     & 105    & 5607  & 1267  & 1529  & 5700  & 0     & 1803  \\
                    & (0.06\%) & (0.66\%) & (35.00\%) & (7.91\%) & (9.54\%) & (35.58\%) & (0.00\%) & (11.25\%) \\ 
    \textbf{Google}     & 2     & 1304   & 10674 & 8410  & 2789  & 5673  & 0     & 11767 \\
                    & (0.00\%) & (3.21\%) & (26.28\%) & (20.70\%) & (6.87\%) & (13.97\%) & (0.00\%) & (28.97\%) \\
    \midrule
    \textbf{US} & & & & & & & & \\
    & & & & & & & & \\
    \textbf{M. Cloud} & 0     & 1448   & 12848 & 10542 & 4516  & 696   & 0     & 3488  \\
                    & (0.00\%) & (4.32\%) & (38.31\%) & (31.43\%) & (13.47\%) & (2.08\%) & (0.00\%) & (10.40\%) \\ 
    \textbf{Google}     & 1     & 2627   & 16359 & 12552 & 3226  & 1839  & 1     & 13627 \\
                    & (0.00\%) & (5.23\%) & (32.57\%) & (24.99\%) & (6.42\%) & (3.66\%) & (0.00\%) & (27.13\%) \\ \bottomrule
    \end{tabular}
    \label{tab:biasscalebycountry}
\end{table}

\newpage
\section{Regression Tables for Popularity and Number of Results}\label{app:decimalsregressiontables}

\begin{table}[!ht]
    \centering
    \caption{Regression model considering all countries and only the Politics and Others sections. Model: OLS Method: Least Squares No. Observations: 2507 R-squared: 0.070 Adj. R-squared: 0.070 F-statistic: 189.9 Prob (F-statistic): 1.08e-41 Log-Likelihood: -13365 AIC: 2.673e+04 BIC: 2.675e+04.}
    \begin{tabular}{lcccccc}
        \toprule
        \textbf{Variable} & \textbf{Coefficient} & \textbf{Std. Error} & \textbf{t} & \textbf{P} & \textbf{95\% Confidence Interval} \\
        \midrule
        Constant & -56.8906 & 5.327 & -10.679 & 0.000 & (-67.337, -46.444) \\
        Open PageRank & 13.9703 & 1.014 & 13.779 & 0.000 & (11.982, 15.958) \\
        \bottomrule
    \end{tabular}
\end{table}

\begin{table}[!ht]
    \centering
    \caption{Regression model considering all countries and sections. Model: OLS Method: Least Squares No. Observations: 4296 R-squared: 0.038 Adj. R-squared: 0.038 F-statistic: 169.1 Prob (F-statistic): 6.01e-38 Log-Likelihood: -27415 AIC: 5.483e+04 BIC: 5.485e+04.}
    \begin{tabular}{lcccccc}
        \toprule
        \textbf{Variable} & \textbf{Coefficient} & \textbf{Std. Error} & \textbf{t} & \textbf{P} & \textbf{95\% Confidence Interval} \\
        \midrule
        Constant & -117.8682 & 11.844 & -9.952 & 0.000 & (-141.089, -94.648) \\
        Open PageRank & 30.0569 & 2.311 & 13.004 & 0.000 & (25.525, 34.588) \\
        \bottomrule
    \end{tabular}
\end{table}

\begin{table}[!ht]
    \centering
    \caption{Regression model for Brazil considering all sections. Model: OLS Method: Least Squares No. Observations: 1226 R-squared: 0.058 Adj. R-squared: 0.057 F-statistic: 75.46 Prob (F-statistic): 1.17e-17 Log-Likelihood: -7994.3 AIC: 1.599e+04 BIC: 1.600e+04.}
    \begin{tabular}{lcccccc}
        \toprule
        \textbf{Variable} & \textbf{Coefficient} & \textbf{Std. Error} & \textbf{t} & \textbf{P} & \textbf{95\% Confidence Interval} \\
        \midrule
        Constant & -144.6259 & 22.141 & -6.532 & 0.000 & (-188.064, -101.187) \\
        Open PageRank & 44.9787 & 5.178 & 8.687 & 0.000 & (34.820, 55.137) \\
        \bottomrule
    \end{tabular}
\end{table}

\begin{table}[!ht]
    \centering
    \caption{Regression model for the UK considering all sections. Model: OLS Method: Least Squares No. Observations: 1478 R-squared: 0.057 Adj. R-squared: 0.056 F-statistic: 89.08 Prob (F-statistic): 1.42e-20 Log-Likelihood: -9310.9 AIC: 1.863e+04 BIC: 1.864e+04.}
    \begin{tabular}{lcccccc}
        \toprule
        \textbf{Variable} & \textbf{Coefficient} & \textbf{Std. Error} & \textbf{t} & \textbf{P} & \textbf{95\% Confidence Interval} \\
        \midrule
        Constant & -200.5584 & 24.403 & -8.219 & 0.000 & (-248.426, -152.690) \\
        Open PageRank & 41.4653 & 4.393 & 9.438 & 0.000 & (32.848, 50.083) \\
        \bottomrule
    \end{tabular}
\end{table}

\begin{table}[!ht]
    \centering
    \caption{Regression model for the US considering all sections. Model: OLS Method: Least Squares No. Observations: 2189 R-squared: 0.070 Adj. R-squared: 0.070 F-statistic: 164.5 Prob (F-statistic): 2.32e-36 Log-Likelihood: -12874 AIC: 2.575e+04 BIC: 2.576e+04.}
    \begin{tabular}{lcccccc}
        \toprule
        \textbf{Variable} & \textbf{Coefficient} & \textbf{Std. Error} & \textbf{t} & \textbf{P} & \textbf{95\% Confidence Interval} \\
        \midrule
        Constant & -151.2867 & 13.710 & -11.035 & 0.000 & (-178.173, -124.400) \\
        Open PageRank & 32.0083 & 2.496 & 12.826 & 0.000 & (27.114, 36.902) \\
        \bottomrule
    \end{tabular}
\end{table}

\newpage
\section{Regression Tables for Lin-PC1 and Number of Results}\label{app:linpc1regressiontables}

\begin{table}[!ht]
    \centering
    \caption{Lin-PC1's effect on the number of results, controlling for Open PageRank. Model: OLS Method: Least Squares No. Observations: 1262 R-squared: 0.139 Adj. R-squared: 0.138 F-statistic: 101.8 Prob (F-statistic): 1.03e-41 Log-Likelihood: -8392.5 AIC: 1.679e+04 BIC: 1.681e+04.}
    \begin{tabular}{lcccccc}
        \toprule
        \textbf{Variable} & \textbf{Coefficient} & \textbf{Std. Error} & \textbf{t} & \textbf{P} & \textbf{95\% Confidence Interval} \\
        \midrule
        Constant & -475.2636 & 45.247 & -10.504 & 0.000 & (-564.031, -386.496) \\
        Open PageRank & 110.1783 & 7.741 & 14.232 & 0.000 & (94.991, 125.366) \\
        Lin-PC1 & -142.9796 & 41.883 & -3.414 & 0.001 & (-225.149, -60.811) \\
        \bottomrule
    \end{tabular}
\end{table}

\begin{table}[!ht]
    \centering
    \caption{Lin-PC1's relationship with Open PageRank scores. Model: OLS Method: Least Squares No. Observations: 1262 R-squared: 0.096 Adj. R-squared: 0.095 F-statistic: 133.1 Prob (F-statistic): 2.41e-29 Log-Likelihood: -1305.6 AIC: 2615 BIC: 2625.}
    \begin{tabular}{lcccccc}
        \toprule
        \textbf{Variable} & \textbf{Coefficient} & \textbf{Std. Error} & \textbf{t} & \textbf{P} & \textbf{95\% Confidence Interval} \\
        \midrule
        Constant & 4.5213 & 0.104 & 43.330 & 0.000 & (4.317, 4.726) \\
        Lin-PC1 & 1.6723 & 0.145 & 11.537 & 0.000 & (1.388, 1.957) \\
        \bottomrule
    \end{tabular}
\end{table}

\begin{table}[!ht]
    \centering
    \caption{Lin-PC1's direct effect on the number of results, showing no statistical significance ($p > 0.05$). Model: OLS Method: Least Squares No. Observations: 1262 R-squared: 0.001 Adj. R-squared: -0.000 F-statistic: 0.9257 Prob (F-statistic): 0.336 Log-Likelihood: -8486.6 AIC: 1.698e+04 BIC: 1.699e+04.}
    \begin{tabular}{lcccccc}
        \toprule
        \textbf{Variable} & \textbf{Coefficient} & \textbf{Std. Error} & \textbf{t} & \textbf{P} & \textbf{95\% Confidence Interval} \\
        \midrule
        Constant & 22.8890 & 30.882 & 0.741 & 0.459 & (-37.697, 83.475) \\
        Lin-PC1 & 41.2762 & 42.900 & 0.962 & 0.336 & (-42.887, 125.440) \\
        \bottomrule
    \end{tabular}
\end{table}

\newpage
\section{Regression Tables for Political Bias and Number of Results}\label{app:politicalbiasregressiontables}

\begin{table}[!ht]
    \centering
    \caption{Regression on the impact of political bias on the number of results an outlet might have, considering only results in the Politics and Others sections and controlling for Open PageRank. Model: OLS Method: Least Squares No. Observations: 913 R-squared: 0.202 Adj. R-squared: 0.201 F-statistic: 115.4 Prob (F-statistic): 2.17e-45 Log-Likelihood: -5074.0 AIC: 1.015e+04 BIC: 1.017e+04.}
    \begin{tabular}{lcccccc}
        \toprule
        \textbf{Variable} & \textbf{Coefficient} & \textbf{Std. Error} & \textbf{t} & \textbf{P} & \textbf{95\% Confidence Interval} \\
        \midrule
        Constant & -253.2822 & 18.324 & -13.823 & 0.000 & (-289.244, -217.321) \\
        Open PageRank & 46.8546 & 3.092 & 15.155 & 0.000 & (40.787, 52.922) \\
        Bias Scale & 2.0302 & 2.302 & 0.882 & 0.378 & (-2.487, 6.547) \\
        \bottomrule
    \end{tabular}
\end{table}

\begin{table}[!ht]
    \centering
    \begin{tabular}{lcccccc}
        \toprule
        \textbf{Variable} & \textbf{Coefficient} & \textbf{Std. Error} & \textbf{t} & \textbf{P} & \textbf{95\% Confidence Interval} \\
        \midrule
        Constant & 6183.7364 & 192.830 & 32.068 & 0.000 & (5805.732, 6561.740) \\
        bias\_scale & 86.4907 & 24.714 & 3.500 & 0.000 & (38.044, 134.938) \\
        Open PageRank & -375.1311 & 27.856 & -13.467 & 0.000 & (-429.738, -320.524) \\
        \bottomrule
    \end{tabular}
    \caption{US: Regression on the ordered scale of political bias (far-left to far-right), where lower values relate to being more positioned to the left. The model is run against the order of the results in Google, where lower means a more prominent position, and controlled for their Open PageRank score. It shows a significant relationship in which a higher political bias scale rating (more to the right) means appearing further down in Google's results (higher order). Model: OLS Method: Least Squares No. Observations: 7064 R-squared: 0.031 Adj. R-squared: 0.030 F-statistic: 111.7 Prob (F-statistic): 1.75e-48 Log-Likelihood: -63724 AIC: 1.275e+05 BIC: 1.275e+05.}
\end{table}

\begin{table}[!ht]
    \centering
    \caption{UK: Analysis did not yield statistically relevant results. Regression on the ordered scale of political bias (far-left to far-right), where lower values relate to being more positioned to the left. The model is run against the order of the results in Google, where lower means a more prominent position, and controlled for their Open PageRank score. This updated analysis indicates a significant relationship where a higher political bias scale rating (more to the right) means appearing further down in Google's results (higher order). Model: OLS Method: Least Squares No. Observations: 3982 R-squared: 0.051 Adj. R-squared: 0.051 F-statistic: 107.3 Prob (F-statistic): 4.13e-46 Log-Likelihood: -33595 AIC: 6.720e+04 BIC: 6.722e+04.}
    \begin{tabular}{lcccccc}
        \toprule
        \textbf{Variable} & \textbf{Coefficient} & \textbf{Std. Error} & \textbf{t} & \textbf{P} & \textbf{95\% Confidence Interval} \\
        \midrule
        Constant & 4141.4909 & 148.340 & 27.919 & 0.000 & (3850.661, 4432.321) \\
        bias\_scale & -18.7711 & 15.840 & -1.185 & 0.236 & (-49.827, 12.285) \\
        decimals & -300.9091 & 20.641 & -14.578 & 0.000 & (-341.378, -260.441) \\
        \bottomrule
    \end{tabular}
\end{table}

\newpage
\section{Regression Tables for Article Age and Order}\label{app:recencyregressiontables}

\begin{table}[!ht]
    \centering
    \caption{Overall: Regression on the impact of article recency (publish time in hours) and popularity (Open PageRank) on weighted order. Model: OLS Method: Least Squares No. Observations: 132106 R-squared: 0.024 Adj. R-squared: 0.024 F-statistic: 1612 Prob (F-statistic): 0.00 Log-Likelihood: -19339 AIC: 3.868e+04 BIC: 3.871e+04.}
    \begin{tabular}{lcccccc}
        \toprule
        \textbf{Variable} & \textbf{Coefficient} & \textbf{Std. Error} & \textbf{t} & \textbf{P} & \textbf{95\% Confidence Interval} \\
        \midrule
        Constant & -0.0137 & 0.004 & -3.234 & 0.001 & (-0.022, -0.005) \\
        Open PageRank & 0.0306 & 0.001 & 45.056 & 0.000 & (0.029, 0.032) \\
        Publish Time (hours) & -0.0020 & 6.41e-05 & -30.711 & 0.000 & (-0.002, -0.002) \\
        \bottomrule
    \end{tabular}
\end{table}

\begin{table}[!ht]
    \centering
    \caption{Brazil: Regression on the impact of article recency (publish time in hours) and popularity (Open PageRank) on weighted order. Model: OLS Method: Least Squares No. Observations: 48581 R-squared: 0.020 Adj. R-squared: 0.020 F-statistic: 488.4 Prob (F-statistic): 9.57e-211 Log-Likelihood: -7482.1 AIC: 1.497e+04 BIC: 1.500e+04.}
    \begin{tabular}{lcccccc}
        \toprule
        \textbf{Variable} & \textbf{Coefficient} & \textbf{Std. Error} & \textbf{t} & \textbf{P} & \textbf{95\% Confidence Interval} \\
        \midrule
        Constant & -0.0359 & 0.009 & -4.092 & 0.000 & (-0.053, -0.019) \\
        Open PageRank & 0.0405 & 0.002 & 24.220 & 0.000 & (0.037, 0.044) \\
        Publish Time (hours) & -0.0020 & 0.000 & -18.391 & 0.000 & (-0.002, -0.002) \\
        \bottomrule
    \end{tabular}
\end{table}

\begin{table}[!ht]
    \centering
    \caption{UK: Regression on the impact of article recency (publish time in hours) and popularity (Open PageRank) on weighted order. Model: OLS Method: Least Squares No. Observations: 36167 R-squared: 0.065 Adj. R-squared: 0.065 F-statistic: 1257 Prob (F-statistic): 0.00 Log-Likelihood: -5284.8 AIC: 1.058e+04 BIC: 1.060e+04.}
    \begin{tabular}{lcccccc}
        \toprule
        \textbf{Variable} & \textbf{Coefficient} & \textbf{Std. Error} & \textbf{t} & \textbf{P} & \textbf{95\% Confidence Interval} \\
        \midrule
        Constant & -0.2121 & 0.009 & -23.518 & 0.000 & (-0.230, -0.194) \\
        Open PageRank & 0.0603 & 0.001 & 45.347 & 0.000 & (0.058, 0.063) \\
        Publish Time (hours) & -0.0021 & 0.000 & -17.551 & 0.000 & (-0.002, -0.002) \\
        \bottomrule
    \end{tabular}
\end{table}

\begin{table}[!ht]
    \centering
    \caption{US: Regression on the impact of article recency (publish time in hours) and popularity (Open PageRank) on weighted order. Model: OLS Method: Least Squares No. Observations: 47358 R-squared: 0.021 Adj. R-squared: 0.021 F-statistic: 510.6 Prob (F-statistic): 4.04e-220 Log-Likelihood: -5892.6 AIC: 1.179e+04 BIC: 1.182e+04.}
    \begin{tabular}{lcccccc}
        \toprule
        \textbf{Variable} & \textbf{Coefficient} & \textbf{Std. Error} & \textbf{t} & \textbf{P} & \textbf{95\% Confidence Interval} \\
        \midrule
        Constant & -0.0421 & 0.008 & -5.124 & 0.000 & (-0.058, -0.026) \\
        Open PageRank & 0.0316 & 0.001 & 25.125 & 0.000 & (0.029, 0.034) \\
        Publish Time (hours) & -0.0018 & 0.000 & -17.287 & 0.000 & (-0.002, -0.002) \\
        \bottomrule
    \end{tabular}
\end{table}

\newpage
\section{Regression Tables for Number of Texts in Media Cloud and Google Results}\label{app:mediacloudregressiontables}

\subsection{Overall Tables}

\begin{table}[!ht]
    \centering
    \caption{Overall: Logistic regression on the impact of the number of texts in Media Cloud's database in the chances of appearing in Google's results for that same query. Model: Logit Method: MLE No. Observations: 15653 Pseudo R-squared: 0.05791 Log-Likelihood: -7457.4}
    \begin{tabular}{lcccccc}
        \toprule
        \textbf{Variable} & \textbf{Coefficient} & \textbf{Std. Error} & \textbf{z} & \textbf{P} & \textbf{[0.025, 0.975]} \\
        \midrule
        Number of Texts & 0.0990 & 0.004 & 26.542 & 0.000 & (0.092, 0.106) \\
        Intercept & -1.8052 & 0.027 & -67.798 & 0.000 & (-1.857, -1.753) \\
        \bottomrule
    \end{tabular}
\end{table}

\begin{table}[!ht]
    \centering
    \caption{Overall: Odds ratios, accuracy, and ROC AUC for the logistic regression model.}
    \begin{tabular}{lcc}
        \toprule
        \textbf{Metric} & \textbf{Value} \\
        \midrule
        Odds Ratio (Number of Texts) & 1.104062 \\
        Odds Ratio (Intercept) & 0.164449 \\
        Accuracy & 0.7989522775186865 \\
        ROC AUC & 0.7210221926426942 \\
        \bottomrule
    \end{tabular}
\end{table}

\begin{table}[!ht]
    \centering
    \caption{Overall: Confusion Matrix for the logistic regression model.}
    \begin{tabular}{l|cc}
        \toprule
        & \textbf{Predicted Negative} & \textbf{Predicted Positive} \\
        \midrule
        \textbf{Actual Negative} & 12258 & 204 \\
       \textbf{ Actual Positive} & 2943 & 248 \\
        \bottomrule
    \end{tabular}
\end{table}

\newpage
\subsection{Country Specific Tables}

\begin{table}[!ht]
    \centering
    \caption{Brazil: Logistic regression on the impact of the number of texts in Media Cloud's database in the chances of appearing in Google's results for that same query. Model: Logit Method: MLE No. Observations: 2585 Pseudo R-squared: 0.04149 Log-Likelihood: -1451.8}
    \begin{tabular}{lcccccc}
        \toprule
        \textbf{Variable} & \textbf{Coefficient} & \textbf{Std. Error} & \textbf{z} & \textbf{P} & \textbf{[0.025, 0.975]} \\
        \midrule
        Number of Texts & 0.0617 & 0.006 & 9.787 & 0.000 & (0.049, 0.074) \\
        Intercept & -1.3116 & 0.056 & -23.403 & 0.000 & (-1.421, -1.202) \\
        \bottomrule
    \end{tabular}
\end{table}

\begin{table}[!ht]
    \centering
    \caption{UK: Logistic regression on the impact of the number of texts in Media Cloud's database in the chances of appearing in Google's results for that same query. Model: Logit Method: MLE No. Observations: 3315 Pseudo R-squared: 0.08451 Log-Likelihood: -1966.5}
    \begin{tabular}{lcccccc}
        \toprule
        \textbf{Variable} & \textbf{Coefficient} & \textbf{Std. Error} & \textbf{z} & \textbf{P} & \textbf{[0.025, 0.975]} \\
        \midrule
        Number of Texts & 0.1429 & 0.009 & 15.919 & 0.000 & (0.125, 0.160) \\
        Intercept & -1.2944 & 0.056 & -23.268 & 0.000 & (-1.403, -1.185) \\
        \bottomrule
    \end{tabular}
\end{table}

\begin{table}[!ht]
    \centering
    \caption{US: Logistic regression on the impact of the number of texts in Media Cloud's database in the chances of appearing in Google's results for that same query. Model: Logit Method: MLE No. Observations: 9753 Pseudo R-squared: 0.04428 Log-Likelihood: -3700.4}
    \begin{tabular}{lcccccc}
        \toprule
        \textbf{Variable} & \textbf{Coefficient} & \textbf{Std. Error} & \textbf{z} & \textbf{P} & \textbf{[0.025, 0.975]} \\
        \midrule
        Number of Texts & 0.0881 & 0.005 & 17.430 & 0.000 & (0.078, 0.098) \\
        Intercept & -2.2131 & 0.038 & -58.750 & 0.000 & (-2.287, -2.139) \\
        \bottomrule
    \end{tabular}
\end{table}

\newpage
\section{Validation Dataset Description and Analysis}\label{app:validationdataset}

An additional dataset with information for 12 additional days, between April 29\textsuperscript{th} 2024 and May 11\textsuperscript{th} 2024, was collected to validate some of the findings of the main analysis. These include the patterns in source concentration and recency, as these do not involve collecting additional data besides that gathered from Google's systems. Overall, the results are quite similar to the data collected originally, providing additional evidence that the findings in this investigation generalize. The data collection used the exact same Python scripts and AWS machine used for the main analysis.

\subsection{Dataset Description}\label{app:validationdatasetdescription}

The additional dataset had a total of a total of \texttt{1,228} \textit{keywords}, rendering a total of \texttt{77,887} \textit{results}. These account for \texttt{28,967} unique \textit{URLs} from \texttt{2,987} \textit{outlets}.

\begin{table}[ht]
    \centering
    \caption{Number of outlets, queries, and URLs for each country;Total URLS (Once per query) ignores repetition of results that appear more than once for the same query (as they were executed multiple times), but will count again if it also appears as a result for a different query.}
    \begin{tabular}{lrrrr}
        \toprule
        & \textbf{Unique Outlets} & \textbf{Unique Queries} & \textbf{Unique URLs} & \textbf{Total URLs} \\
        \textbf{Country} &  &  &  & (\textit{once per query}) \\
        \midrule
        BR & 896 & 474 & 10165 & 11567 \\
        UK & 1003 & 332 & 8662 & 10036 \\
        US & 1522 & 422 & 11029 & 12987 \\
        \bottomrule
    \end{tabular}
\end{table}
    
\subsection{Concentration}\label{app:validationdatasetconcentration}

Similarly to the main dataset, the validation data showed low average HHI, \textsc{0.09}, and Gini \textsc{0.33}, when considering individual queries. Considering the data horizontally (across multiple queries), the same pattern of concentration is seen. The low HHI, \textsc{0.005}, and high Gini, \texttt{0.799}, again indicate the behavior of a select few outlets dominating the results, while the vast majority compete for similar small shares.

This translates into 2.85\% of the outlets accounting for 50\% of the results, or 1.1\% if considering their weighted rankings. That concentration is also shown by the Kolmogorov-Smirnov statistic (\(D = 0.894\), \(p < 0.001\)) and the \( \text{CR}_k \) values in tables~\ref{tab:crkvalidation} and~\ref{tab:crkweightedvalidation}.

\begin{table}[ht]
    \centering
    \caption{\( \text{CR}_k \) shows the concentration of Top \(k\) outlets for each country.}\label{tab:crkvalidation}
    \begin{tabular}{lrrrrr}
        \toprule
        \textbf{Country} & \textbf{\( \text{CR}_4 \)} & \textbf{\( \text{CR}_8 \)} & \textbf{\( \text{CR}_{10} \)} & \textbf{Results} & \textbf{News Sources} \\
        \midrule
        BR & 0.175155 & 0.273110 & 0.306538 & 29197 & 896 \\
        UK & 0.200280 & 0.303560 & 0.346698 & 20701 & 1003 \\
        US & 0.098789 & 0.169602 & 0.200222 & 27989 & 1522 \\
        \bottomrule
    \end{tabular}
\end{table}

\begin{table}[ht]
    \centering
    \caption{\( \text{CR}_k \) shows the concentration of Top \(k\) outlets for each country, considering the weighted score for each entry. This places more emphasis on records the higher they appear on the results page.}\label{tab:crkweightedvalidation}
    \begin{tabular}{lrrrrr}
        \toprule
        \textbf{Country} & \textbf{\( \text{CR}_4 \)} & \textbf{\( \text{CR}_8 \)} & \textbf{\( \text{CR}_{10} \)} & \textbf{Results} & \textbf{News Sources} \\
        \midrule
        BR & 0.324691 & 0.649142 & 0.810528 & 29197 & 896 \\
        UK & 0.320757 & 0.641515 & 0.800638 & 20701 & 1003 \\
        US & 0.301547 & 0.603094 & 0.752831 & 27989 & 1522 \\
        \bottomrule
    \end{tabular}
\end{table}

\subsection{Variation and Recency}\label{app:validationdatasetvariation}

The same patterns of gradually updating the results found in the original analysis were spotted in the validation dataset, seen in Jaccard Index and RBO values being lower between first and last iterations than they are between sequential ones. The progression in average article age suggests that content is being replaced for morue recent texts. If that was not the case, they should increase at an average rate of three hours, which was the time interval between collections. 

\begin{table}[ht]
    \centering
    \caption{Descriptive statistics of the Jaccard Index and RBO comparing first iteration to last.}
    \begin{tabular}{lccccc}
    \toprule
    \textbf{Statistic} & \textbf{Count} & \textbf{Mean} & \textbf{Std. Deviation} & \textbf{Minimum} & \textbf{Maximum} \\
    \midrule
    Jaccard & 1228 & 0.469 & 0.192 & 0.000 & 1.000 \\
    RBO & 1228 & 0.286 & 0.357 & 0.000 & 0.999 \\
    \bottomrule
    \end{tabular}
    \label{tab:jaccardanalysisvalidate}
\end{table}

\begin{table}[ht]
    \centering
    \caption{Average Jaccard, RBO, and time difference between article publishing and data collection (Article Age) over different iterations. Jaccard and RBO scores represent a comparison with the iteration that came immediately before. The mean variations in the columns were tested using ANOVA. They are all statistically significant (Jaccard: \(p < 0.001\), RBO: \(p < 0.05\))}
    \begin{tabular}{lccc}
    \toprule
    \textbf{Scraping Iteration} & \textbf{Avg. Jaccard} & \textbf{Avg. RBO} & \textbf{Avg. Article Age (Hours)} \\
    \midrule
    1 &      &      & 13.44 \\
    2 & 0.49 & 0.51 & 14.49 \\
    3 & 0.51 & 0.52 & 15.40 \\
    4 & 0.53 & 0.54 & 16.56 \\
    5 & 0.54 & 0.56 & 17.10 \\
    \bottomrule
    \end{tabular}
    \label{tab:analysisoveriterationsvalidate}
\end{table}

\end{appendices}

\end{document}